\documentclass[useAMS,usenatbib,referee]{mn2e}
\usepackage{graphicx}
\usepackage{natbib}
\usepackage{times}
\usepackage{float}
\usepackage{amsmath}
\usepackage{epsfig}
\usepackage{xcolor}
\definecolor{blue}{cmyk}{0,0,0,1}
\definecolor{white}{cmyk}{0,0,0,0}
\definecolor{majenta}{rgb}{1,0,1}
\definecolor{majenta}{cmyk}{0,0,0,1}
\newcommand{\bmth}[1]{\mbox{\boldmath${#1}$}}

\newcommand{\bsf}[1]{\mbox{\boldmath$\mathsf{#1}$}}
\begin{document}

\title 
[Tidal interaction near the Kraft break] 
{The tidal interaction of an orbiting giant planet  with  a star near the Kraft break:  the  excitation of $r$-modes and the retention 
of orbital and spin angular momenta misalignment} 
\author[J. C. B. Papaloizou and  G.J. Savonije ]
{ J. C. B. Papaloizou $^{1}$\thanks{E-mail: J.C.B.Papaloizou@cam.ac.uk (JCBP)}, 
{G.J. Savonije$^{2}$\thanks{E-mail: G.J.Savonije@uva.nl (GJS)},}
	    \\
	$^{1}$ DAMTP, Centre for Mathematical Sciences, University of
           Cambridge, Wilberforce Road, Cambridge CB3 0WA 	 \\    
	$^{2}$Anton Pannekoek Institute of Astronomy, University of Amsterdam, Science Park 904,
	NL-1098 XH, Amsterdam }

\maketitle

\date{Accepted. Received; in original form}

\pagerange{\pageref{firstpage}--\pageref{lastpage}} \pubyear{2010}


\label{firstpage}

{\textcolor{blue}{\begin{abstract}
\textcolor{majenta}{			
 In this paper we extend the previous work of Papaloizou \& Savonije on tidal interactions
 between a solar mass star and a closely orbiting giant planet which is such that  the orbital and stellar spin angular { momentum  
  directions}  are  misaligned.  
 Here we consider the situation when the central star 
  {has a mass of}  $1.3 M_{\odot}$ and {is}  in the vicinity of the Kraft break.
We find and determine the properties of the lowest order $r$ modes and the tidal response arising from the 
secular non axisymmetric forcing associated with a misaligned orbit. 
We find that the response of the thin convective envelope, as well as the shift of $r$ mode frequencies from the low rotation frequency,  limit can be  
understood  by adopting a vertically averaged model that is similar to the well known one  governed by the Laplace tidal equation for an incompressible ocean. 
 From our results we are able to estimate lower bounds on realignment time scales 
for hot Jupiter  systems with orbital periods in the range $2.8-5 d$ and rotation periods in the range $5-31 d$ that indicate the process is indeed markedly less effective 
than for a solar type star. This is  on account of  there being less dissipation in a relatively smaller convective envelope as well as  the generally faster rotation
and hence larger spin angular momentum 
 expected for the more massive star.
  }
\end{abstract}}
  
  \begin{keywords}
hydrodynamics - celestial mechanics -
planet - star interactions - stars: 
rotation - stars: oscillations (including pulsations )- stars:solar-type
\end{keywords}

\section{Introduction}\label{intro}
 \textcolor{blue}{ We extend the  studies of the tidal interaction of a solar mass primary with a Jupiter
 mass secondary in a close circular orbit carried out by  \citet {PS23} (hereafter PS) and \citet{PS24} (hereafter PS1)  to 
 consider the case  when the central star is instead a higher mass star of mass $1.3M_{\odot}$, with effective temperature $T_{eff} = 6401K,$  thus being in the neighbourhood
 of the Kraft break. More details of the model are given in table \ref{table1}.  As in PS1 we allow  the  stellar spin and orbital angular momentum vectors to be misaligned.}

  \textcolor{blue}{ 
 The initial distribution  of alignment angles of close orbiting giant planets may arise from the action of a number of processes  including  disc migration \citep[eg][]{LP86} and dynamical interactions  within a multi-planet system or with a stellar companion  \citep[see e.g.][]{Siegel, Wrightetal, Wuetal}.
    While disc migration is expected to be associated with alignment, the other mechanisms   may produce significant primordial misalignment of the angular momentum vectors in the system. }
    
 \textcolor{blue}{   For planets close to the central star, subsequent tidal interaction can result in  orbital  evolution. 
 For an isolated hot Jupiter in an orbit, constrained to be circular on account of tides acting on the planet, this could lead to  a combination of  orbital decay  and a tendency to  alignment of the orbital and spin angular momenta. 
  Dissipation  resulting from the action of turbulent viscosity  \citep[see eg.][]{Zahn1977,Duguid2020}  in the stellar convective envelope  and the excitation of inwardly propagating $g$ modes may play  an important role \citep[see e.g.][]{O2014}.
The extent to which this evolution occurs is important for relating  parameters of observed systems to conditions just post formation. }

\textcolor{majenta} { \cite{Albrecht12} and \cite{Lai12} argued that the effective components of the tidal forcing potential,  which are stationary in the inertial frame and have azimuthal mode numbers $n=\pm 1,$ only occur in non-aligned systems.  In a frame co-rotating  with the star, assumed to rotate uniformly,
they  have forcing frequencies in the inertial range and potentially excite inertial waves in the convective envelope with relatively strong dissipation rates. Being stationary in the inertial frame this type of tidal forcing  will not lead to orbital decay in hot Jupiter systems.}

\textcolor{majenta}{ In a frame that is  co-rotating   with the  central  star and aligned with its rotation axis, the magnitude of the  forcing frequencies associated with the above secular terms become $\Omega_s$ for $|n|=1,$ and $2 \, \Omega_s$  for $|n|=2$, with $2\pi/\Omega_s$ being the stellar rotation  period.
   Accordingly the tidal interaction may be associated both with strong inertial mode responses in the convective envelope \citep[e.g.][]{PP81, OL2007, IP2010,  O2014}  and r mode   responses  in the radiative interior of the central star \citep[e.g.][]{PP78, Dewberry2023}.
 We remark that  the discussion of \citet{Attia} indicates some  influence of tides on the alignments of hot Jupiters, in the direction  of stronger alignments  being  favoured for cooler stars with extended convective envelopes though the effect seems to be modest. 
This also follows from our calculations presented here together with those in PS1.}

\textcolor{majenta} {	In addition we locate the low order free   $r$ mode frequencies  for degrees  $l'=1,$ and $l'=3$ 
and determine the resonance widths  of the lowest order modes. 
 We determine  the tidal response associated with the secular forcing terms that arise in the misaligned case finding that, as for the solar mass case,
 the narrow resonance widths effectively lead to a non resonant response.
	We use our results to provide  lower bounds  for spin orbit alignment  time scales for hot Jupiter  systems with this type of central star referring to some observed systems listed in table \ref{table0}.}

\begin{table} 	
	\begin{center}
		{\color{blue}
			\begin{tabular} {cccccccccc} 
				\hline 	
				$M/M_{\odot}$&$R_s/R_{\odot}$&$L/L_{\odot}$& $T_{eff} (K)$&
				$I (gm cm^2)$&$\Omega_c (s^{-1})$&$M_{convc}/M$&$R_{ce}/R_s$&$M_{conve}/M$& age Gy.\\
				$1.30 $& $ 1.4768 $& $3.25$ &$ 6.4011\times10^3$&.  
				$ 1.16\times 10^{54} $&  $3.987\times 10^{-4}$&  $ 5.73\times 10^{-2}  $&0.883 & $2.19\times 10^{-4}$&$1.20$\\		
				\hline
		\end{tabular}}
	\end{center}  
	\caption{ \textcolor{blue}{Parameters for the stellar model: The first column gives the mass in solar masses, the second and third give the radius and luminosity  in solar units respectively, the fourth the effective temperature,
			the fifth and sixth the moment of inertia and critical angular velocity in c.g.s. units respectively, the seventh gives the mass fraction in the convective core, and  the eighth and ninth  the  fractional radius  of the inner
			convective envelope boundary and the convective envelope mass fraction respectively. The last column gives the age defined as the time to evolve from the initial state 
			with the uniform hydrogen mass fraction, $X=0.7,$ to when the central value reaches $X_c=0.4.$
	} }
	\label{table1}
\end{table}

\textcolor{majenta} {The plan of this paper is as follows:
	In Section \ref{tidalresponse} we outline the procedure used for obtaining the tidal response to forcing  by  potential perturbations with angular  dependence proportional to a spherical  harmonic  of degree $l=2$, 	as defined in a coordinate system co-rotating with the central star with the $Z$ axis coincident with its rotation axis.
	This forcing is  assumed to arise from a perturber in a circular orbit with angular momentum that has a  general  non-zero  inclination with respect to the stellar spin angular momentum \citep[see also][]{IP2021}.  In this section we also describe how the  resulting tidal dissipation is calculated. In Section \ref{visc} we give the expression used to calculate the dynamic viscosity in the convective envelope.}
		 
		 \textcolor{majenta}{In Section \ref{thinconvaveraging} we  summarise the equations governing  the response of a simplified  thin convective envelope model based on vertical averaging.   
		 More details are given in appendices \ref{AS1}-\ref{AS4}.
		 As outlined  in appendix \ref{Pertt1a} we use this model to estimate the departure of the $r$ mode frequencies from
		 their values in the zero frequency limit.  This  is found to be consistent  with our numerical work in Section \ref{l1nr0resonanceshift}. 
		 This  shift accounts for why  $r$ modes are effectively
		 non resonant, given their very narrow half-widths that we determined for rotation periods of interest.}
		 
\textcolor{majenta}	{The numerical results are given  in Section \ref{NumRes}.  After reviewing  the main properties of $r$ modes and their eigen frequency determination in Section \ref{rmoderes}, we go on to give results for $r$ modes with $l'=1$ 
	and  $l'=3$ in Sections \ref{l'=1} and  \ref{l'=3}, respectively.}

 We go on to  find the non resonant  tidal response of our model for rotation periods in the range $5-28$$d$ in Section \ref{Nonresresults}.
 We focus on responses to   tidal forcing that appears stationary in an inertial frame. In a frame co-rotating with the central star 
 the forcing frequencies are,  $-\Omega_s,$ and $-2\Omega_s,$ leading to the possibility of a strong interaction with inertial modes in the convective envelope  that occurs only when the spin and orbital angular momenta are misaligned \citep[eg.][]{Lai12}.}
 
 Assuming  this interaction  was the dominant one leading to alignment, with changes to the magnitude of the orbital angular momentum being negligible, 
 we discuss the effects on tidal evolution  in Sections \ref{Effecttidalevol}. In particular we  present expressions, based on the earlier work of PS1, for the rate of evolution of the spin orbit angle under the assumption 
 of fixed orbital angular momentum  in  Sections   \ref{spinorbevol} and \ref{m0evol}.

  


\textcolor{majenta}
{In Sections \ref{2.8days} and \ref{shortperiod} we consider the application to  hot Jupiter systems with orbital periods in the range $0.94-5$ days listed in table \ref{table0}. 
This indicates very little alignment of misaligned  systems with non decaying orbits  should occur in their lifetimes. The tidal  process is less effective  than for systems with  a solar type star on account of both weaker dissipation for forcing of comparable strength in a smaller convective envelope, as well as  shorter rotation periods and thus the relatively larger spin angular momentum expected  for more massive models. 
This important  feature, which necessarily makes attaining alignment more difficult,  is related to the expected ineffectiveness of stellar winds \citep[eg.][]{Skum72} beyond the Kraft break.}
 Finally. we summarise and discuss our results in Section \ref{discuss}. 
 
 \begin{table} 	
 	\begin{center}
 		\textcolor{majenta}{
 			\begin{tabular} {|c|c|c|c|c|c|c|c|c|} 
 				\hline 	
 				&System$\hspace{3mm}$ &$M_{p}/M_J\hspace{3mm}$ & $P_{orb}\hspace{1mm} d\hspace{3mm}$ &$M_s/M_{\odot}\hspace{3mm}$&$R_s/R_{\odot}\hspace{3mm}$& $T_{eff} \hspace{1mm}K\hspace{3mm}$&
 				$V\sin i \hspace{1mm}kms^{-1}\hspace{3mm}$&Alignment\\    
 				&&&&&&&\\
 				&WASP-7& $ 0.96\pm 0.3$& $4.95$ &  $1.36^{+0.26}_{-0.19} $ & $1.47^{ +0.06}_{-0.07}$   &$ 6562^{+118}_{-119}$&$17\pm 2$&N\\
 				\hline
 				&HAT-P-$30\hspace{2mm}$& $ 0.83\pm 0.18$& $2.81$ &  $1.25^{+0.25}_{-0.15} $ & $1.34 \pm 0.06$   &$ 6338^{+162}_{-124}$&$2.2\pm 0.5$ &N\\
 				\hline
 				&HAT-P-$9\hspace{2mm}$& $ 0.75^{+0.64}_{-0.63} $ & $3.92$ & $1.28 {\pm 0.07}$  &$1.3^{+0.068}_{-0.073} $   &$ 6340^{+73}_{-56}$&$11.9\pm 1 $&Y\\
 				\hline
 				&Kepler-$8\hspace{2mm}$& $ 0.59^{+0.13}_{-0.12} $& $3.52$ & $1.46^{+0.09}_{-0.18}$  &$1.46^{+0.09}_{-0.18} $   &$ 6346^{+435}_{-153}$&$10.5\pm 0.7 $&Y\\
 				\hline
 				&Kelt-$4\hspace{1mm}$& $ 0.90 \pm 0.06 $& $2.99$ & $1.20^{+0.07}_{-0.06}$ &$1.60\pm 0.04$&$ 6207\pm 75$&$5.8\pm 0.45$&N\\
 				\hline	
 				&&&&Short \hspace{1mm} Period&\hspace{-1mm}Systems&&\\
 				\hline
 				&WASP-12& $ 1.47\pm 0.08$& $1.09$ &  $1.35\pm 0.14 $ & $1.57\pm 0.07$   &$ 6360^{+200}_{-100}$&$2.2\pm 1.5$&N\\
 				\hline
 				&WASP-18$\hspace{0mm}$& $ 10.2\pm 0.35$& $0.94$ &  $1.29\pm 0.06 $ & $1.32\pm 0.06$   &$ 6432\pm 48$&$11.0\pm 1.5$&Y\\
 				\hline
 		\end{tabular}}
 	\end{center}
 	\caption{ \textcolor{majenta}{Parameters of exoplanet systems with central stars near the Kraft break which have orbiting Hot Jupiters.
 			The first column is the system name, the second column is the planet mass in Jupiter masses, the third column is the orbital period in days, the fourth column is
 			the mass of the central star in solar masses, the fifth column is the radius of the central star in solar radii, and the sixth column is the effective temperature.
 			The seventh column is $V\sin i$ for the central star and the eight column indicates the degree of alignment between the stellar spin and orbital angular momenta
 			with N indicating  significant misalignment and Y smaller departures from alignment $< \sim 30^{\circ}.$ The data in this table was obtained from
 			the NASA  Exoplanet Archive (see references therein and  \citet{Akeson2013}. }}
 	\label{table0}
 \end{table}

 {\section{Calculation of the tidal response and energy dissipation} \label{tidalresponse}
\textcolor{blue}	
{ Following PS and PS1 we use the fact that the  tidal response  of the primary can be  assembled by summing individual responses ($m=0,\pm 1,\pm 2$ ; $n=\pm 1,\pm 2$ ) to the real parts of harmonically varying tidal potentials in the non rotating stellar frame of the form 
 \begin{equation}
	{\cal U}_{n,m} =\frac{ r^2}{2} c_{tid,m}Y_{2,n}(\theta, \phi) \, \exp(-{\rm i}m\left(n_o t + \varpi+\gamma \right)) \label{potpert1}
\end{equation}
\begin{equation}
c_{tid,m}  =-  \frac{8 \pi GM_p}{5 a^3}  Y_{2,m}(\pi/2, 0).\hspace{4mm}  
\label{jpe711e}
\end{equation}
 We adopt $M_p = 9.543 \times 10^{-4} M_{\odot}$ being equal to Jupiter's mass.
The spherical harmonic of degree $l$ and order $n$ is denoted by $Y_{l,n}(\theta,\phi)$ whereas $(r,\theta,\phi)$ are spherical polar coordinates with origin at the centre of mass of the primary, the $Z$ axis pointing in the direction of the spin angular momentum ${\bf S}$. This is as defined the stellar frame (see PS1).
  We denote the orbital frequency in the non rotating frame with origin at the primary, the orbital frame,  by $n_o$. The longitude of periastron, which would play a role once the eccentricity is non vanishing,  is $\varpi.$ }
  
\textcolor{blue} { The  Wigner factor associated with the transformation  between spherical harmonics defined in the orbital frame  and those defined in the stellar frame 
\citep[see][and PS1]{IP2021}
has been omitted in (\ref{potpert1}). This  depends on the angle $\beta$ between the spin angular momentum  $\bf S$ along the $Z$-axis and orbital angular momentum $\bf L$ along the $Z'$-axis as defined in the orbital frame. 
 This factor  can be  included later when required  (see discussion in Section 2.3 of PS1). This is necessary
  in order to perform tidal evolution calculations as in PS1 and below.
The angle $\gamma$ that also appears in the Wigner transformation is 
  \footnote{The angle $\gamma$ is  in fact ignorable in the response calculation and may be set to  zero}
  taken to be zero. }

 Importantly the total angular momentum, ${\bf J}={\bf L}+{\bf S},$ is conserved by the tidal interaction.
We set $L= |{\bf L}|,$  $S= |{\bf S}|,$  $J= |{\bf J}|,$  and denote the angle between ${\bf L}$ and  ${\bf J}$ by $i.$
The Lagrangian displacement associated with the response to the perturbing potential ${\cal U}_{n,m}$ is $\mbox{{\boldmath$\xi$}}_{n,m}  \exp(-{\rm i}m\left(n_o t + \varpi+\gamma \right))$
The associated Eulerian density perturbation  is  $\rho^\prime_{n,m} \exp(-{\rm i}m\left(n_o t + \varpi+\gamma \right))$ with similar expressions for the other perturbed state variables.

\textcolor{majenta}{
From this we calculate the overlap integral
\begin{align}
\hspace{-5.2cm} Q_{n,m}= \int_V {\rho'_{n,m}({\bf r})}{r^2}Y_{2,n}^*(\theta,\phi) dV\hspace{5mm} 
\label{overlap}
\end{align}
\color{blue}
This quantity can be related  to the mean rate of change of the  kinetic energy associated with forcing due to  the real part of the  corresponding potential given by (\ref{potpert1}) (see PS1).} Thus
\begin{align}
\frac{dE_{kin,n,m}}{dt} = -\frac{c_{tid,m} \omega_{f,n,m}}{4} Im (Q_{n,m}). \label{Edot},
 \end{align}
 where,  $Im,$ denotes that the imaginary part is to be taken.\\ 
 This   is expected to be negative definite for a primary with stable free oscillations. Evaluation of these quantities for values of $n$ and $m$  in the interval $(-2,2),$  aided by the relation
\begin{align}
\frac{dE_{kin,-n,-m}}{dt} =\frac{dE_{kin,n,m}}{dt}
\label{dotER}
\end{align}
\textcolor{blue}{
enables determination of the rate of evolution of the semi-major axis, $a,$ and $\beta$, see section \ref{Effecttidalevol}.}

\textcolor{blue}{ The dominant energy dissipation occurs by the turbulent viscosity in the convective envelope of the $1.3 \, M_\odot$ star. The rate of kinetic energy decrease  is numerically calculated by applying the viscous stress tensor for compressible flow, see equation(49) in PS1.  The associated rate of radiative damping $\cal{D}$
for prescribed $(n,m)$ in the radiative core follows as (see PS)
\begin{align}
\cal{D} \,= \rm{Im}	\int \rm{i} \, \left(	\Gamma_3 -1 \right) \, \left(\nabla \cdot \bmth{\xi}_{n,m}^*\right) \, \left(\nabla \cdot {\bmth{F}'_{n,m}}\right) \, dV
\end{align}
where $\bmth{F}'_{n,m}  \exp(-{\rm i}m\left(n_o t + \varpi+\gamma \right))$
 is the radiative flux perturbation.}

\subsection{Turbulent viscosity}\label{visc}
	  As found  for  solar type stars, the tidal torque on the $1.3M_{\odot}$ star  is directly related to the viscous dissipation 
 in the convective envelope. 
As also found by PS for $1M_{\odot},$  radiative dissipation in the radiative regions of the star is  found to be significantly smaller and can be neglected. 
 As in PS and PS1,  the contribution of the artificial viscous damping in the radiative core, introduced  to  deal with potentially unresolved  short wavelength gravity waves,    
  is  not included in  the stated viscous dissipation rates on account of its resolution dependence.

As in   PS a turbulent kinematic viscosity of the form  
 $\nu(r)$  was taken from \cite{Duguid2020} as
\begin{equation}
	\nu(r)= 
	\frac{ {\frac{1}{3}}
	{ \mathcal{L}_{mx} v_{c} }}
	{(1+({\tau_{c}}/{P_{osc}})^{s})}
	\label{turbvisc}
\end{equation}
was assumed.
Here the  mixing length $ \mathcal{L}_{mx}= \alpha |H_P|$ is scaled by the parameter $\alpha=2$.
 The local pressure scale height $H_P$  and the local convective velocity $v_{c}(r)$ are taken from the  MESA input stellar  model.   The 
 characteristic time associated with  the turbulent convection is taken to be \big($\tau_c=1/\sqrt{|N^2|}$\big), where $|N|.$  the Brunt-V\"{a}is\"{a}l\"{a} frequency
 corresponds to the growth rate of convective modes.  The factor  $(\tau_c/P_{osc})^s,$ with  $P_{osc}=2 \pi/\omega_{f,n,m}$
  in the denominator with  $s=2$ provides a viscosity reduction  that is expected when there is a mismatch between the forcing frequency and  estimated convective time 
  scale \citep[see][]{Duguid2020}.  We stress the well known significant  uncertainties  involved in this type of viscosity specification.

We introduced a thin layer with artificial viscosity in a transition layer between the radiative core and the inner boundary of the convective envelope at $r/R_s= r_{cb}/R_s=0.883$ by extrapolating inwards from  the kinematic viscosity $\nu_{cb}$ at this boundary  according to \\
$\nu(r) = \nu_{cb} \exp \{-\left((r_{cb}-r)/(c_{*} H_P)\right)^2\} $
 down to the adopted minimum artificial viscosity of \\  $\nu_{min}$ in the radiative core.  Here  $H_P$ is  evaluated 
  at the convective boundary and the scaling factor is 
 $c_{*}$. For all calculations isolating $r$ mode resonances we adopted $\nu_{min}=10^3$ and $c_*=0.5.$ 
  For non resonant  forcing  with $n=-1, m=0$, the adopted minimum core viscosity  was  $\nu_{min}~=~10^9$ cgs. 
 For non resonant forcing with $(n=-2, m=0)$ and $(n=-2,m=-2)$
   the larger value  $\nu_{min}=10^{11}$ cgs  was required to damp short wavelength grid oscillations  of the displacement vector in the core.
\textcolor{blue}   
 { For all non resonant calculations we adopted $c_{*}=0.05.$}
We remark that for this model, for largest value of $\nu_{min}$ adopted, $\nu_{min}/(R_s^2 \, \Omega_c)\sim 2.34\times10^{-8}.$ Thus for the smallest angular velocity 
considered with this value of $\nu_{min},$  namely, $\Omega_s\sim 0.006 \,\Omega_c,$ 
 the Ekman number $\nu_{min}/(R_s^2\Omega_s)\sim 3.9\times10^{-6}.$ 

\subsection{ A thin convective envelope model based on vertical averaging }\label{thinconvaveraging}

We have developed a simplified model to describe the linear response of the convective envelope, approximated as barotropic and inviscid,  when its relative thickness to radius is small.
This is based on an approach similar to that used to derive the Laplace  tidal equations \citep[e.g.][]{LH} for an incompressible ocean.
Details and definitions of notation are given in appendices \ref{AS1}-\ref{AS4}. In Section \ref{AS4}    a governing equation (\ref{B4X})  for  the enthalpy perturbation ${\cal W}' = P'/\rho,$  which
in the limit of zero centrifugal distortion used here reads\\
\begin{align}
	-\frac{ \rho_{Rce}R_{ce}^2\sigma^2}{\Sigma}\left(   \bmth {\xi}_{R_s}-\bmth{\xi}_{Rce}\right )\cdot{\hat{\bf r}}   
	= {\cal L}(W')
	\label{B4Xtext}
\end{align}
We remark that ${\cal W}',$  and hence from equations (\ref{B21}) of appendix \ref{AS4},
the horizontal components of the displacement, $\xi_{\theta}$ and $\xi_{\phi},$
are  functions of $\theta,$ and $\phi$ alone.
In addition $ \rho_{Rce}$ is the unperturbed density at the lower boundary of the convective envelope, 
$\bmth{\xi}_{R_s}, \bmth{\xi}_{Rce},$ and $\xi_{r,Rce}$ denote the Lagrangian displacement at the surface radius, the Lagrangian displacement
on the lower boundary of the convective envelope,
and its radial component there respectively.
The surface density of the convective envelope is $\Sigma,$
and we take $\sigma$ to be the oscillation frequency which here may be either free or forced.
In the latter case $\sigma \equiv \omega_{f,n,m}.$
The self-adjoint operator ${\cal L}({\cal W}')$  is specified  in the limit of zero centrifugal distortion
by equation (\ref{C0}) of  appendix \ref{HM}.\\
It is such that the eigenvalue problem ${\cal L}({\cal W}')= -\lambda{\cal W}'$ defines Hough eigenvalues, $\lambda(\sigma),$ and  associated eigenfunctions. 
These define the angular dependence of $\xi_r$ in the traditional approximation \citep[see e.g.][PS]{SP97}  making a connection between
that and  the application of the vertically averaged thin layer approximation considered here. }

\textcolor{majenta}{ As indicated in appendix \ref{AS4}, equation (\ref{B4Xtext}) may be written in the alternative form
\begin{align}
	-\frac{ \rho_{Rce}R_{ce}^2\sigma^2}{g\Sigma}\left({\cal W}' - \Phi'- g\xi_{r,Rce}\right) 
	= {\cal L}(W'),
	\label{B4atext}
\end{align}
where  $\Phi'$ is the the gravitational potential perturbation which becomes the perturbing tidal potential when self-gravity is neglected and the gravitational acceleration, $g,$
is evaluated at the stellar surface.
The displacement at the lower boundary, $\xi_{r,Rce},$  is determined by the response of the inner radiative zone to  $W'$ and $\Phi'$  there.
These  effectively provide an inner  boundary  condition  for this response. 
As described in Section \ref{Pertt1a}  we write\\  $\xi_{r,Rce}\equiv\xi_{r,Rce}({\cal W}',\Phi',\sigma)= \xi_{r,Rce}({\cal W}',0,\sigma)+\xi_{r,Rce}(0,\Phi',\sigma)$ 
as a linear operator acting on $W'$ and $\Phi'.$ Thus (\ref{B4atext}) becomes
\begin{align}
	-\frac{ \rho_{Rce}R_{ce}^2\sigma^2}{g\Sigma}\left({\cal W}' - \Phi'- g\xi_{r,Rce}(W',\Phi',\sigma)\right)= {\cal L}(W')
	\label{B4antext}
\end{align}
	We make use of the above results and others given in appendices \ref{AS1} - \ref{Tidalresponse} when discussing our numerical solutions below.}

\section{Numerical Results} \label{NumRes} 

\subsection{$r$ mode resonances}\label{detrmodes}  \label{rmoderes}
\textcolor{majenta}  
{We present results for a $1.3 M_{\odot}$ main sequence model obtained with the MESA stellar evolution code version r22.11.1 \citep{Paxton2015}. The parameters of this model are given in table \ref{table1}. For this model we investigated $r$ mode resonances which have resonance frequencies  near
\begin{align}
	\sigma_0 = \frac{2 \, n \,\Omega_s}{l'(l'+1)} \label{rmodfreq}
\end{align}
with the relative deviation $\rightarrow 0$ as $\Omega_s\rightarrow 0$  \citep{PP78}.
 Here $l'$ is an integer equal to $l \pm 1$ where  the modes are isolated  through resonant forcing by a tidal potential  with angular dependence  $\propto Y_{l,n}(\theta,\phi)$
 ( see equation (\ref{potpert1})).}

\textcolor{majenta} {In order to find $r$ mode resonances generated by a Jupiter mass planet in orbit about the star we adopt a  particular  value for the stellar spin frequency $\Omega_s$ while to facilitate  comparison of  tidal forcing calculations with different forcing frequencies, including  non resonant ones, 
we artificially keep $a/R_s$ fixed at the arbitrary value, $a/R_s= 10,$  which corresponds to a fixed orbital period of 5.763 $d$.}
 
 \textcolor{majenta}{Adopting a forcing frequency $\omega_f = \sigma_0 + \epsilon$ whereby $|\epsilon|$ is  small ($ << |\sigma_0|$) we solve (see PS) for the tidal response which is
 determined as a function of  $	\omega_f$.  The r-mode resonance frequency $\omega_0$ is found by applying Brent's iterative method \citep{NumRec96} to search for the forcing frequency which corresponds with a significant maximum in the  kinetic energy associated with the tidal response  (see PS, PS1). Because the tidal solution requires the inversion of a large matrix for each forcing frequency and since it requires typically 15 iterations this process is  time consuming. 
   The actual value of the semi-major axis $a$ corresponding to the resonant frequency and the scaled tidal amplitudes, being proportional at resonance  to $a^{-3}$, follow by solving equation (\ref{forcfreq}) below for $n_o$ (see also equation (\ref{potpert1})) after substituting the numerically calculated resonant frequency $\omega_0$  and using Kepler's law. 
\begin{align}
	\omega_f(n,m) \equiv -m n_o+ n \Omega_s = \omega_0  \label {forcfreq}
\end{align}
For the most important quadrupole tidal forcing with $l=2,$  $r$ modes with $l'=1$ and $l'=3$ are expected to be excited (PS,PS1). Accordingly we focus on these values.}

\subsection{Results for $r$ modes with $l'=1$}\label{l'=1}
\textcolor{majenta}{
For $l'=1$ we must have $|n|=1$, accordingly without loss of generality  we  adopt a forcing potential ${\cal U}_{n,m}$, see equation (\ref{potpert1}),
with   $ n=-1$ and $m=-2.$
It follows from equation (\ref{rmodfreq}) that the resonance frequency is  expected to be close to $-\Omega_s,$  the value corresponding to the rigid tilt mode (see PS1).
Accordingly. from equation (\ref{forcfreq}), the precise location of a resonance  peak corresponds to a small value of $n_o$ which is found by the method decribed in section \ref{detrmodes}.}

\begin{figure}
	\hspace{-0.9cm}	
	\includegraphics[height=9cm, width=9cm]
	{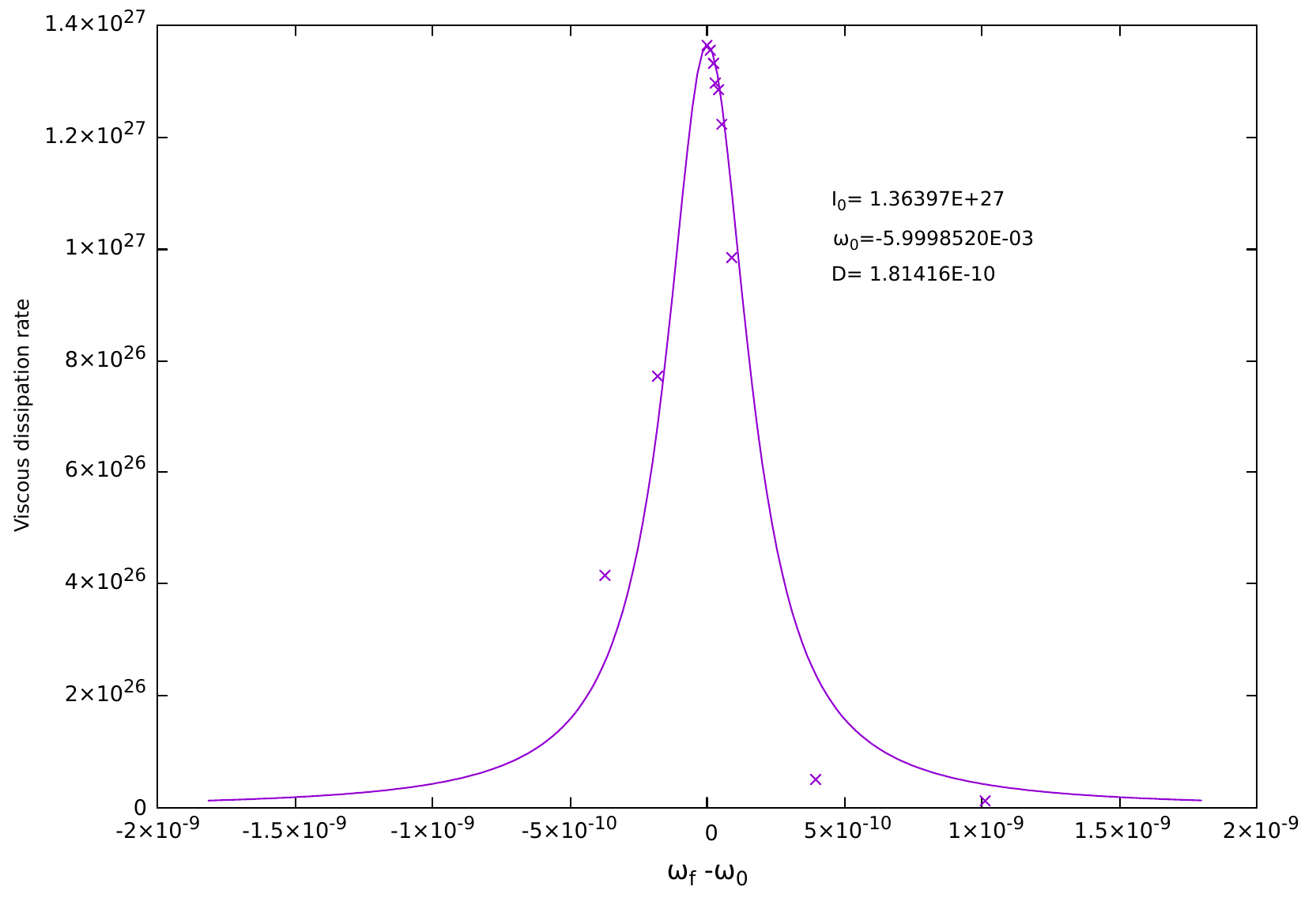}
	\includegraphics[  height=9cm,width=9cm]
	{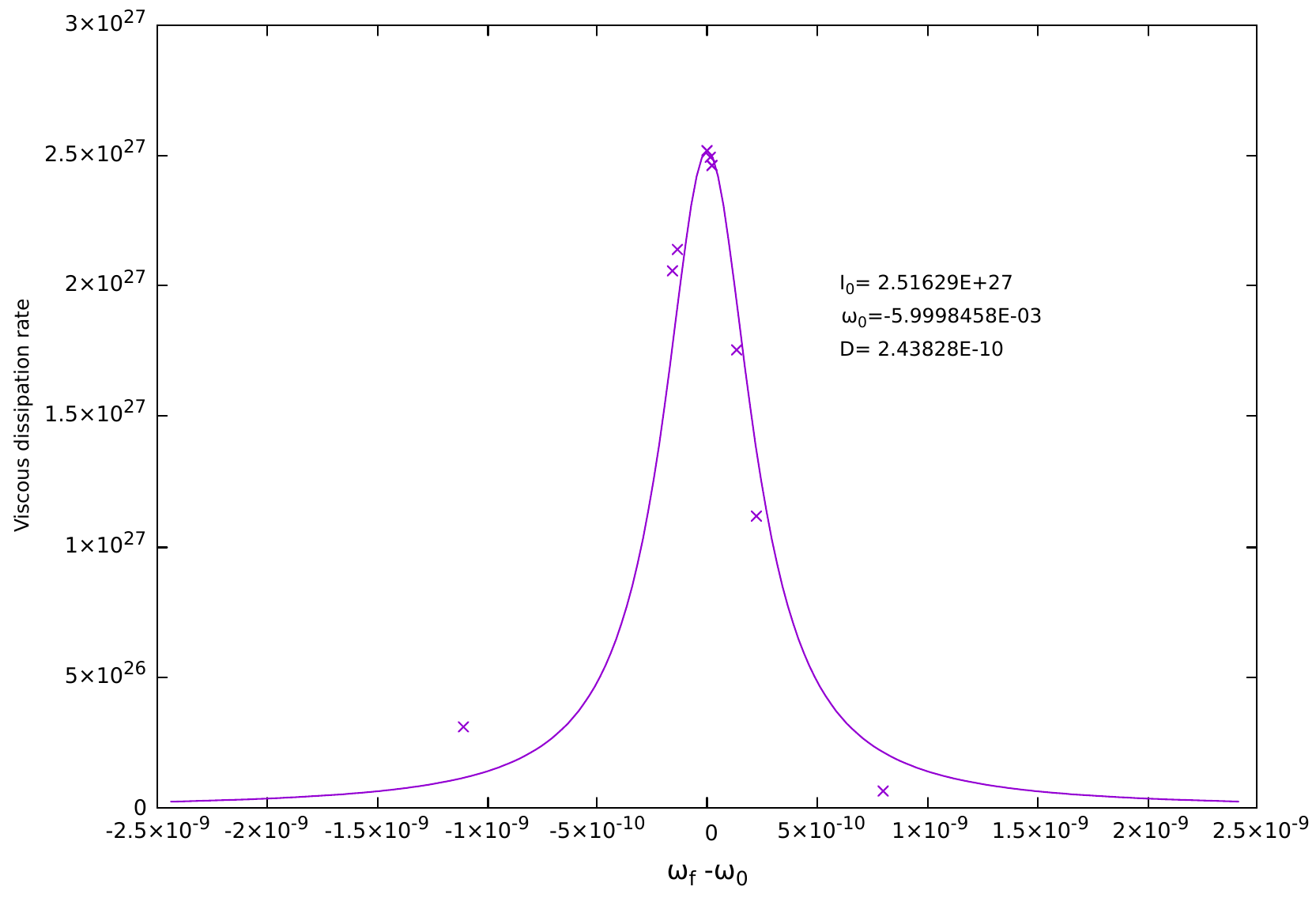}
	\caption{The { left}  panel shows the  resonance curve (with resonance frequency $\omega_0$ and full resonance half width, $D$ in units of $\Omega_c$ associated with the kinetic energy dissipation rate (erg/s) for the $n_r=0$ $r$-mode with $l'=1.$ and $n=-1.$   
		This was located by finding the response to the component of the forcing potential,  ${\cal U}_{-1,-2},$ with $n=-1$ and $m=-2.$ The { right}  panel shows the corresponding resonance curve obtained for the mode with $n_r=1.$  For both cases the value of $\Omega_s$ adopted was $6.00\times 10^{-3} \, \Omega_c.$
		Note that the subscripts $n,$ and $m$ have been removed from
		the forcing frequency $\omega_f.$  }  \label{l1nr0}
\end{figure}

\begin{figure}
	\hspace*{-2cm}
	\includegraphics[angle=0,width=20cm]
	{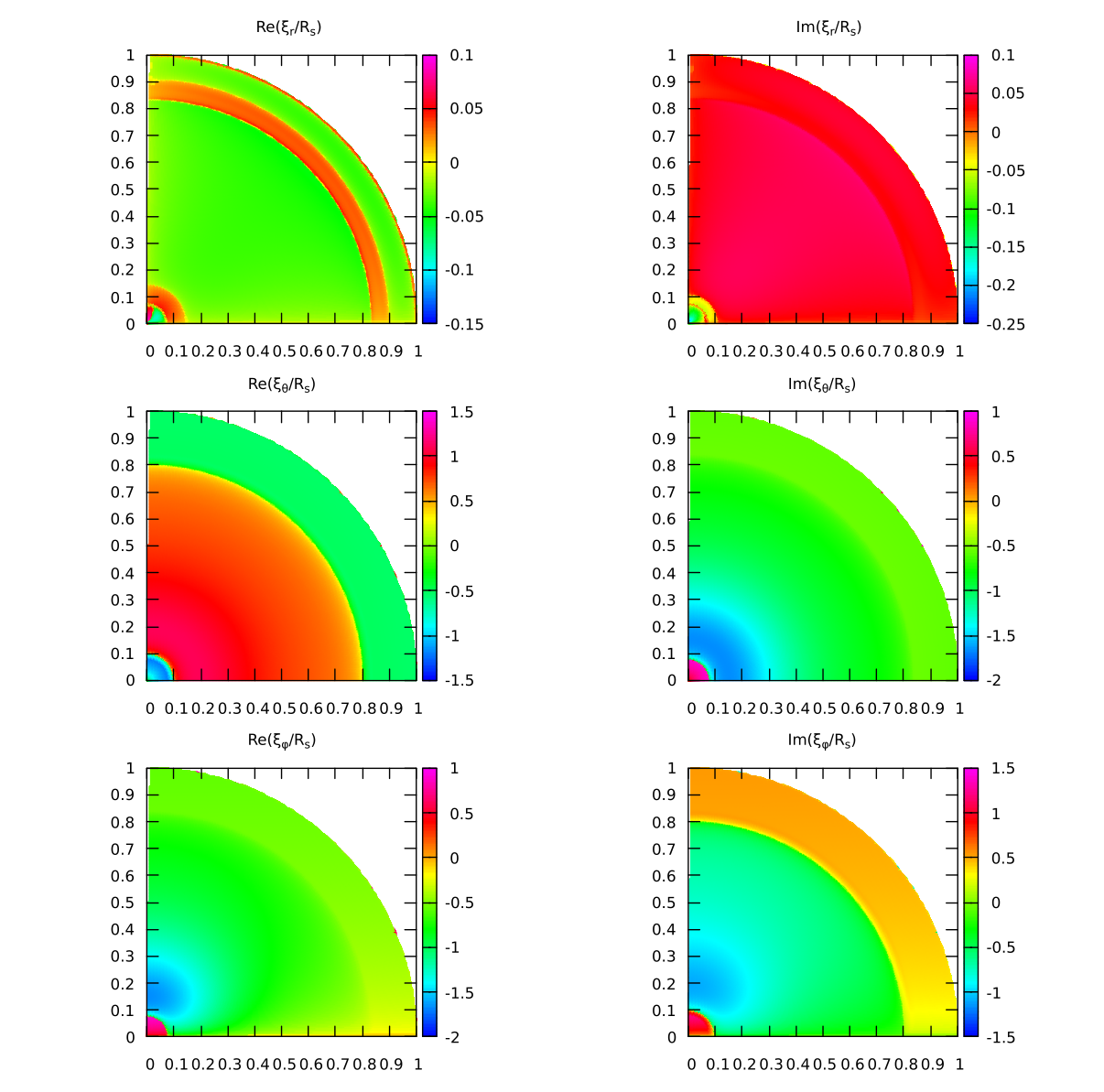}
	\textcolor{majenta} {
		\caption{Contour plots  in the primary's meridional plane $\phi=0$ for the  resonant $l'$=1 $r$ mode, at the  resonance frequency $\omega_0=-5.9998520 \times 10^{-3} \,\Omega_c$, see Fig. \ref{l1nr0}. The adopted stellar spin frequency is $\Omega_s=6.00\times10^{-3}\Omega_{c}$. The resonance frequency is the closest one to $-\Omega_s$. 
			The resonance was obtained by forcing with ${\cal U}_{-1,-2}$  with $(n, m)=(-1,-2)$. The forcing frequency  $\omega_{f,-1,-2}= 2 n_o-\Omega_s $ is very close to $-\Omega_s$ so that $n_0\ll\Omega_s.$
			The Cartesian coordinates along the two axes indicate the relative radius $r/R_s$. 
			The vertical colour bars on the right indicate the local value of  sign($|\xi_x|^{\frac{1}{4}},\xi_x)$, where  $\xi_x$ is  the  component of the displacement vector illustrated
			in units of $R_s$.\label{l1nr0contour} } }
\end{figure}

\textcolor{majenta} {Adopting $\Omega_s = 6.0 \times 10^{-3} \,  \Omega_c$, corresponding to $P_{orb}=30.4 \, d$, a maximum in the response kinetic energy is found for a forcing frequency  $\omega_{f,-1,-2}\equiv \omega_f = \omega_0 =-5.999852 \times 10^{-3} \, \Omega_c$. 
  The very narrow profile of the  total  rate of viscous dissipation in the convective envelope associated with the (unscaled) tidal response  as a function of forcing frequency is illustrated in the left hand panel of  Fig.\ref{l1nr0}. 
  Corresponding} contour plots for the components of the  Lagrangian displacement
 at the resonance frequency are  shown in Fig. \ref{l1nr0contour}. 
 From these plots it is apparent that the angular (but not the radial)  dependences of the horizontal components resemble 
 what is  expected for the rigid tilt mode. These are given by $\xi_{\theta} =Q_0/(R_{ce}\Omega_s^2),$ $Q_0$ being constant, and  $\xi_{\phi}= - {\rm i}\xi_{\theta}\cos\theta$ (see appendix \ref{RTmode}).
 The mode corresponding to this resonance was such that the angular components of the Lagrangian  displacement had the least number of radial nodes. Accordingly we describe this as having $n_r=0$,  even though sign changes of the components of the Lagrangian displacement 
 occur when comparing   their form in the radiative interior with that in both the convective envelope and the convective core.

\begin{figure}
	\hspace{-1.0cm}	
	\hspace*{-1.4cm}	\includegraphics[angle=0,width=20cm] 
	{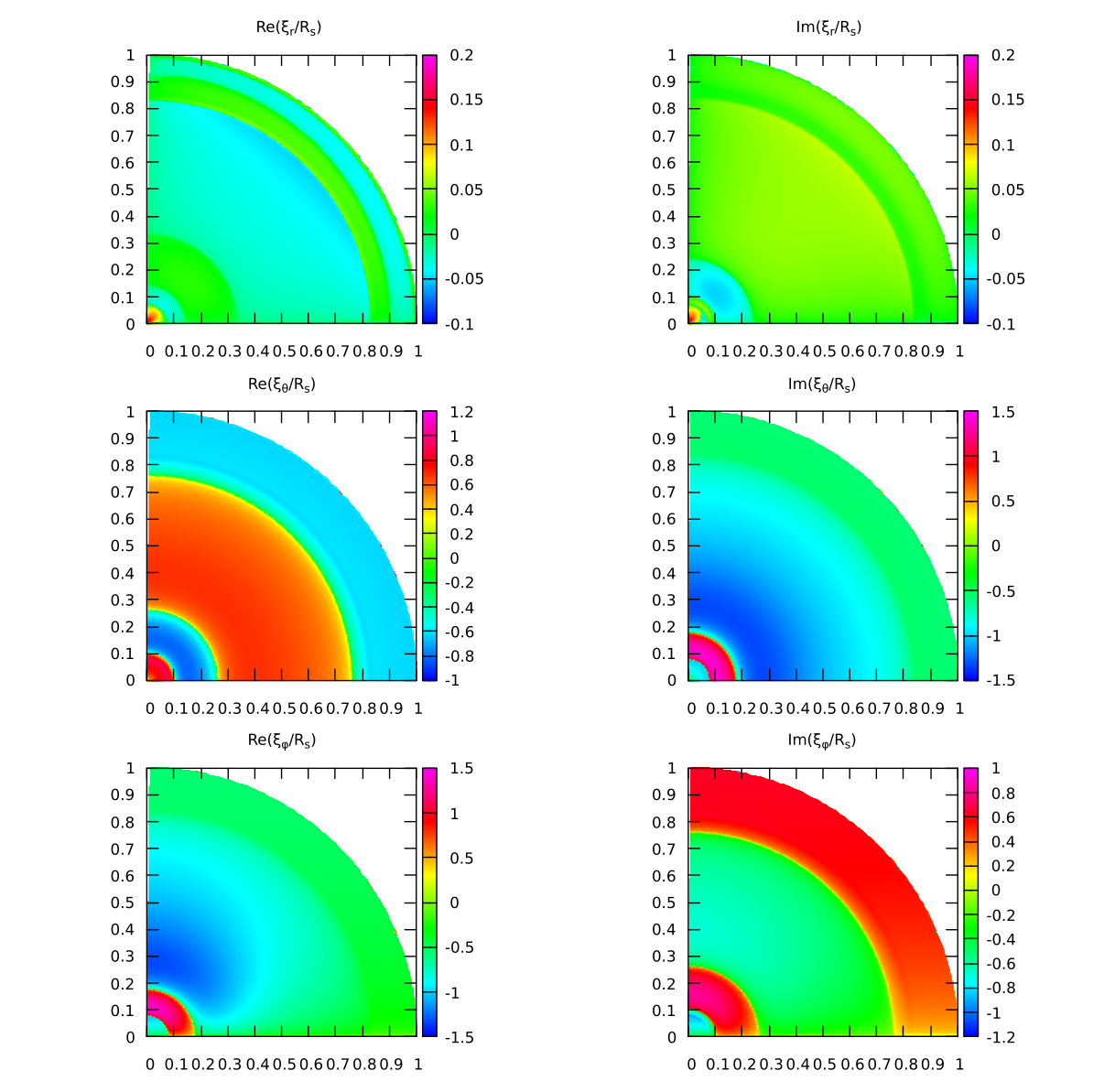}
	\caption{
		\color{majenta} {
			As for Fig. \ref{l1nr0contour} but for the $l'=1$ $r$ mode with $n_r=1$.
			The   resonance frequency  is $\omega_0=-5.9998458 \times 10^{-3} \,\Omega_c.$
			Compared to the case with $n_r=0,$ for the horizontal components of the displacement there is an extra radial node in the inner radiative region.}  
		\label{l1nr1contour} }
\end{figure}

 \textcolor{majenta}{We also located the resonance for which the angular dependence of the horizontal components of the Lagrangian displacement
 has an extra node in the radial direction.  We label the corresponding mode as having  $n_r=1.$
   The resonance frequency in this case was given by $\omega_{f,-1,-2}\equiv \omega_f = \omega_0=-5.9998458 \times 10^{-3}\, \Omega_c.$
  \textcolor{majenta}{
  The total  rate of viscous dissipation in the convective envelope  associated with the tidal response in the neighbourhood of the resonance frequency is shown in the right hand panel}
 of Fig. \ref{l1nr0}.  The corresponding contour plots for the Lagrangian displacement are shown in Fig. \ref{l1nr1contour}. }
  
 We also note that  as  indicated  by  equation (\ref{C3}) of Section \ref{rmodes}, our results are in accordance with the thin vertically averaged model of 
 the convective envelope for which  ${\cal W}'$  $\rightarrow W_0,$   the Hough eigenfunction for eigenvalue $\lambda=0,$ and  $l'=-n=1,$ this having angular dependence
  $\propto \sin\theta\cos\theta.$ 
 As expected from the discussion in appendix \ref{AS4}, 
 the same apparently holds for the angular dependence of $\xi_r$ (see equation (\ref{B1}) for both these modes.  
 

 \textcolor{majenta}{We remark that features seen in the contour plots shown in Figs. \ref{l1nr0contour} and \ref{l1nr1contour},
  as well as those found in other resonant and non resonant responses
 described below,  are  consistent with expectations from the vertically averaged
 thin convective envelope model described in section(\ref{thinconvaveraging}) and in appendices \ref{AS1}-\ref{AS4}. In particular the  
 horizontal displacement in the convective envelope 
 {is}  to a good approximation independent of, $r,$ as expected. 
 In addition the magnitude of $\xi_r$ is smaller than { the magnitude of}  the horizontal component {of the displacement}
 by a factor of order $ 10^{-5} -10^{-4}.$  As explained in appendix \ref{AS4}, this is expected  to be of order $\epsilon _2= \Omega_s^2/\Omega_c^2\sim 4\times 10^{-5}$ for free $r$ modes of oscillation.}
 
{   We remark that for the stellar  model considered here,  the ratio of the thickness to radius of the convective envelope $\epsilon_{1} =0.12$ which is assumed small
when deriving the vertically averaged model. Reasonable agreement with the numerical results is obtained as indicated above.
However, for the solar model considered in PS1 for which $\epsilon_{1} = 0.27$  this is not the case.
 }

\textcolor{majenta}{As noted in PS1, there is no rigid tilt mode associated with our model and so the horizontal displacements are not simply $\propto r$
and $\xi_r\ne 0.$ However the response in the convective envelope is very like a rigid tilt mode with a small non zero $\xi_r.$
As explained  in appendix \ref{RTmode} for the convective envelope to move as if participating in a rigid tilt mode,
 the effect of the self-gravity perturbation resulting almost entirely  from the tilt of the inner regions must be included.
 If the inner regions remain stationary or move differently,   the self-gravity perturbation required for the rigid tilt mode, has to be  replaced by a pressure perturbation
which in turn produces  a  non zero $\xi_r.$ In this way the convective envelope can undergo an approximate rigid tilt while layers below do not participate.
Furthermore the small value of $|\xi_r|$ and other corrections will be associated with only  a relatively small dissipation with the tilt component not contributing.
This tends to make dissipative  tidal responses to low frequency forcing with $|n|=1$ ineffective (see below)}

\textcolor{majenta}{As in PS and  PS1  over a narrow range of forcing frequency, the resonance curves shown in Fig.\ref{l1nr0} match reasonably well
 to  a curve which takes the the  form  $ I_0/
(1+((\omega_f - \omega_0)/D)^2),$ The  parameters $I_0, \omega_0, $ and $D$ that we adopted are indicated for each of the curves illustrated  in Fig. \ref{l1nr0}.}
We see that for the $n_r=0$ mode,  the separation of the frequency at the resonance peak  from $-\Omega_s,$
 the expected value in the limit $\Omega_s\rightarrow 0$ (see PS),
being $1.480\times 10^{-7} \, \Omega_c$ is small.  Even so, this separation greatly exceeds the 
 half-width  $ D= 1.81416\times 10^{-10} \,\Omega_c.$
 The same holds for the $n_r=1$ mode for which the frequency separation from $\Omega_s$ is $1.542\times 10^{-7} \,\Omega_c$ \textcolor{majenta}{
 and $D=2.43828\times10^{-10} \,\Omega_c.$ }The small frequency separation of $r$ modes with successive values of $n_r$  when  compared to their  frequency shift  from the zero frequency limit  was also  found in the aligned case by PS.
 Given their extremely  small half-widths this has the consequence that the tidal response at the forcing frequency $\-\Omega_s$
 that may be significant for the tidal evolution of misalignment is effectively non resonant.
 This was found to be the case for the $1M_{\odot}$ model in PS1.

\textcolor{majenta}{\subsubsection{Estimating the resonant frequency shift from $-\Omega_s$ for $n_r=0$}\label{l1nr0resonanceshift}
The governing equation for normal modes is equation (\ref{B4antext}). We use this to find the  resonant frequency shift from $-\Omega_s$
employing perturbation theory. 
In the first instance we consider\\ $l'=-n=1.$}

\textcolor{majenta}{As  indicated in appendices \ref{AS1}-\ref {AS4}, the factor, $ \rho_{Rce}R_{ce}^2\sigma^2/(g\Sigma),$  on the left hand side is taken to be a small parameter for our perturbation approach,
which at lowest order can be taken to be zero. Then equation (\ref{B4antext}) is simply  ${\cal L}(W') =0$ and
  the Hough eigenvalue, $\lambda,$ is zero corresponding to $\sigma=\sigma_0=2n\Omega_s/(l'(l'+1)$ ( see appendices \ref{HM} and  \ref{rmodes}).
  We \textcolor{majenta}{then} have  ${\cal W}'={\cal W}_0$ as given by (\ref{C3}). We proceed to find the correction to $\sigma$ when the left hand side of (\ref{B4antext})
is restored using perturbation theory.}

\textcolor{majenta}{Doing this we set  $\sigma=\sigma_0+\delta\sigma,$ 
and the eigenvalue  becomes $\delta\lambda_{l',n}.$
This is specified as a function of $\sigma$ by equation (\ref{C4}) in appendix \ref{Pertt}.
In addition the application of perturbation theory to (\ref{B4antext}) as described in appendix \ref{Pertt1a} gives (see equation (\ref{D1}))
\begin{equation}
\frac{ \rho_{Rce}R_{s}^2\sigma_0^2}{g\Sigma}\int^{\pi}_0\sin\theta {\cal W}_0\left( {\cal W}_0 -g\xi_{r,Rce}({\cal W}_0 ,0, \sigma)\right)d\theta
= \delta\lambda_{l',n}(\sigma) \int^{\pi}_0{\cal W}_0^2\sin\theta d\theta.  \label{D1text}    
\end{equation}
Note that the full value of $\sigma =\sigma_0+\delta\sigma,$  rather than just $\sigma_0,$ is retained in  
 the expression, $\xi_{r,Rce}({\cal W}_0,0,\sigma)$. 
We argue that the effect of this term  is   expected to be small in magnitude. However, in the context of the present  numerical calculation,
where it is governed by the response of the interior, its  magnitude  may vary rapidly with $\sigma,$ e.g. exhibiting strong peaks at characteristic values, 
 such that when (\ref{D1text})
is taken more generally  to be an equation for $\sigma,$ 
 a series of  modes of different radial order $n_r$  can be found ( see also PS).  To fully  specify  these, this term will then be significant.
 In view of this, one can  see that the higher order modes may be regarded as contributing to the frequency correction for the $n_r=0$ mode.
 This is not considered  quantitatively in our simple analytic  model but we remark that  the effect is expected to be sensitive to the amount of 
 dissipation associated with the evaluation of  $\xi_{r,Rce}({\cal W}_0,0,\sigma).$ 
}


\begin{figure}
	\hspace{-0.9cm}	
	\includegraphics[height=9cm, width=9cm]
	{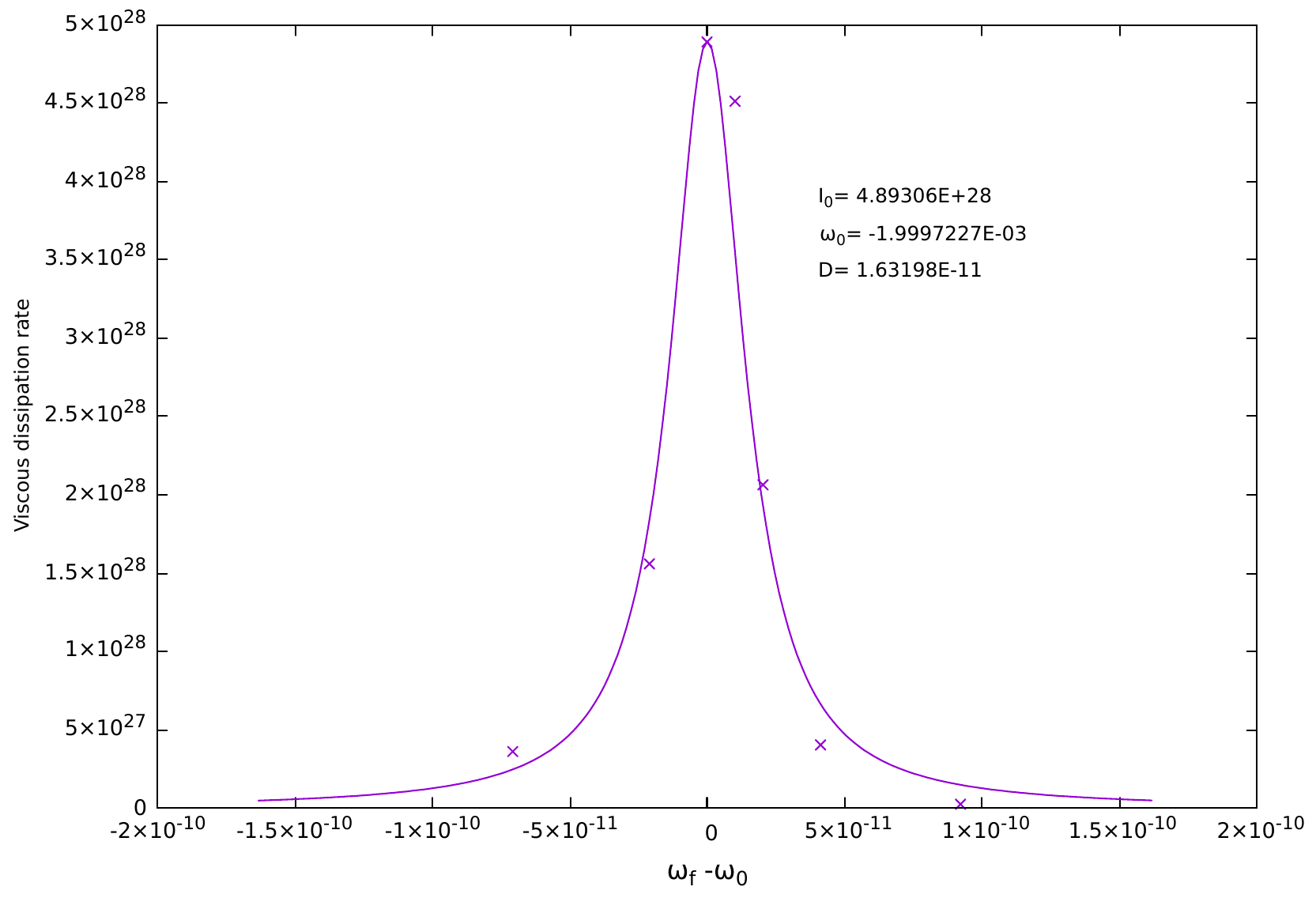}
	\includegraphics[  height=9cm,width=9cm]
	{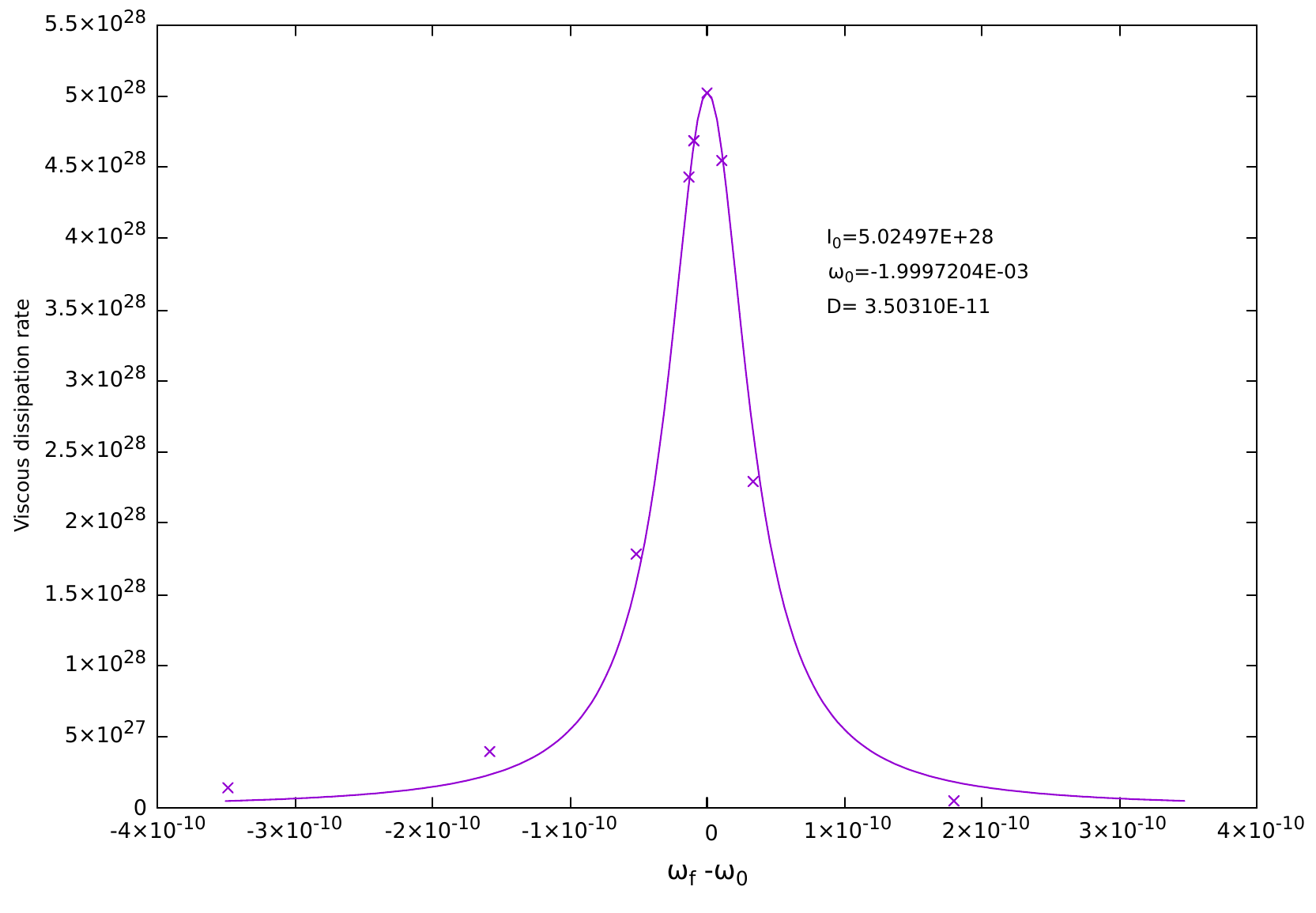}
	\caption
		{The left panel shows the  resonance curve (with resonance frequency $\omega_0$ and full resonance half width, $D$ in units of $\Omega_c$ associated with the kinetic energy dissipation rate (erg/s) for the $n_r=0$ $r$-mode with $l'=3$ and $n=-2.$   
		This was located by finding the response to the component of the forcing potential,  ${\cal U}_{-2,-2},$ with $n=-2$ and $m=-2.$ The right panel shows the corresponding resonance curve obtained for the mode with $n_r=1$. For both cases the value of $\Omega_s$ adopted was $6.00\times 10^{-3} \, \Omega_c.$
		Note that the subscripts $n,$ and $m$ have been removed from the forcing frequency $\omega_f.$  } \label{l3nr0}
\end{figure} 

\textcolor{majenta}{Here we crudely  estimate $\delta\sigma$ neglecting $\xi_{r,Rce}$. This corresponds to all the possible modes with different, $n_r,$ 
 coalescing into a single   mode that is unaffected by the response of the inner radiative region.
In this case equation (\ref{D1text}) with the help of equation (\ref{C4}) gives 
\begin{align}
\frac{ \rho_{Rce}R_{ce}^2\sigma_0^2}{g\Sigma}=\delta\lambda\equiv \delta\lambda_{1,-1}= -\frac{10\delta\sigma}{\sigma_0}.
\end{align}
Thus
\begin{align}
\frac{\sigma-\sigma_0}{\Omega_c} \equiv \frac{\omega_0+\Omega_s}{\Omega_c}= \frac{\Omega_s^3R_s}{4\Omega_c^3 h_{ce} }\left(\frac{R_{ce}}{R_s}\right)^2,\label{shift}
\end{align}
where $h_{ce}$ is the depth of the convection zone and we adopted $\Sigma=2\rho_{Rce}h_{ce}/5.$
 For the calculation illustrated in Fig. \ref{l1nr0contour} we find that (\ref{shift}) gives $\omega_0+\Omega_s= 3.5\times 10^{-7} \,\Omega_c,$ which
 in view of the approximations adopted is in reasonable agreement with the numerical result. } 
 
 \textcolor{majenta}{ Recall that the frequency differences between the $n_r=0$ mode and modes with  $n_r > 0$  are small and within the context of this model
 are  associated with different contributions arising  from  $\xi_{r,Rce}({\cal W}_0,0,\sigma).$}

\textcolor{majenta}{\subsection{Results for $r$ modes with $l'=3$}\label{l'=3}
	We also  located $r$ modes for $l'=3$  corresponding to $n_r=0,$ and  $n_r=1$ for the  model of the $1.3 M_{\odot}$ star  with $\omega_s= \Omega_s/\Omega_c=6 .00\times 10^{-3} \Omega_c.$ 
	 As for $l'=1,$ the tidal forcing was due to a Jupiter mass planet with $a/R_s= 10.$ 
	For modes with  $l'=3$ we adopt a forcing potential,  ${\cal U}_{n,m},$ with $ n=-2$ and  $m=-2.$  This corresponds to the predominant quadrupole tidal forcing.
	A response to this with $l'=3$  is produced 
	through the action of  the Coriolis force.
	Recalling (\ref{rmodfreq}) which gives  $r$ mode frequencies in the limit $\Omega_s\rightarrow 0,$ we have 
	$ \omega_{f,n,m}= 
	2n\Omega_s/l'(l'+1).$
	For $l'=3,$  and $n=-2$,   this gives 
	$ \omega_{f,-2,-2}=\sigma_0=  -\Omega_s/3.$}
	
	\textcolor{majenta} {As for the $l'=1$ case,   the orbital frequency $n_0$ was adjusted to provide a required resonant
	forcing frequency through the relation, $\omega_{f,n,m}  = - m n_o ~+~ n \Omega_s .$ The response  can then be scaled 
	to correspond to this  value. }
\textcolor{majenta}{In Fig.\ref{l3nr0} we show  the total rate of viscous dissipation associated with the tidal responses in the neighbourhood of the resonance frequencies
	of the modes  with $n_r=0$ (left panel) and  $n_r=1$ (right panel).
	The determined resonance frequencies were $\omega_{f,-2,-2}\equiv \omega_f=  -1.9997227\times 10^{-3} \,\Omega_c,$ and
	$\omega_{f,-2,-2}\equiv \omega_f=  -1.9997204\times 10^{-3} \,\Omega_c$ respectively. }\\

As for the case $l'=1$ we can attempt to account for the  separation of the resonant frequency of the mode with $n_r=0$ from $\sigma_0.$
	We make use of  (\ref{D1text}) for $l'=3, n=-2$ while noting that ${\cal W}'_0$ then becomes the Hough eigenfunction 
	as specified through equation (\ref{C3}) in appendix \ref{rmodes} with $l'=3,n=-2.$ While from appendix \ref{Pertt} the discussion below  equation (\ref{C4}) leads to\\
	$\delta\lambda=-
	0.93\delta\sigma/\sigma_0 \equiv \delta\lambda_{3,-2}(\sigma).$
	Thus, following a parallel discussion to that given for  the $l'=1$ case, from equation (\ref{D1text}) we find
	\begin{align}
		\frac{\sigma-\sigma_0}{\Omega_c} \equiv \frac{\omega_0+\Omega_s/3}{\Omega_c}= \frac{0.1\Omega_s^3R_s}{\Omega_c^3 h_{ce} } \left(\frac{R_{ce}}{R_s}\right)^2,\label{shift3}
	\end{align}
	This gives a value of   $1.4\times 10^{-7}$
	 which compares to the calculated value of $2.8\times 10^{-7}.$
	As for $l'=1,$ these are in reasonable agreement considering the approximations involved.
	As for  $l'=1,$ the frequency difference between the frequencies associated with the $n_r=0$ and those with larger values of $n_r$  is significantly less than the shift from $\sigma_0,$ and a similar discussion applies.

\textcolor{majenta}{ Contour plots for the components of the Lagrangian displacement for the response at the resonance peak  for $n_r=0$ are shown in 
	Fig.\ref{l3nr0contour} and those for the mode with $n_r=1$ in Fig. \ref{l3nr1contour}. 
	\begin{figure}
	\hspace{-1.0cm}	
	\hspace*{-1.3cm}	\includegraphics[angle=0,width=20cm] 
	{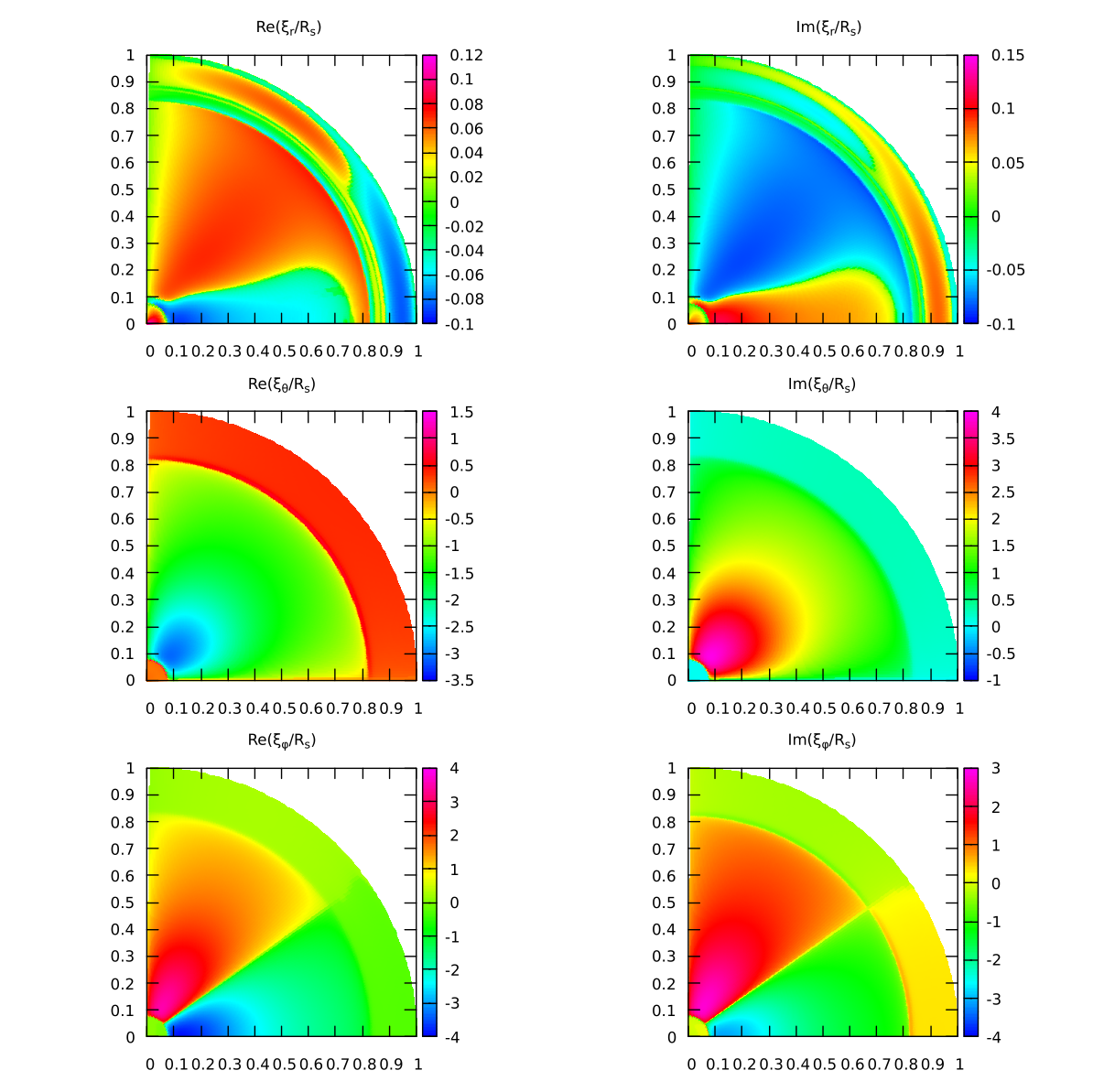}
	\caption{Contour plots  in the primary's meridional plane $\phi=0$ for the  $l'$=3 $r$ mode with $n_r=0.$ 
		\color{majenta} The  resonance frequency is $\omega_0=-1.9997227 \times 10^{-3} \,\Omega_c$  this being the closest one to $-\Omega_s/3.$
	This was obtained by resonant forcing with, ${\cal U}_{-2,-2},$  with $(n, m)=(-2,-2) ,$  and $\Omega_s=6.00\times10^{-3}\Omega_{c}$ { (see Fig. \ref{l3nr0}).}
	The forcing frequency  $\omega_{f,-2,-2}= 2 n_o-2\Omega_s \equiv \omega_0$ is very close to $-\Omega_s/3$ so that $n_0\sim 5\Omega_s/6.$   
	The Cartesian coordinates along the two axes indicate the relative radius $r/R_s$. 
	The vertical colour bars on the right indicate the local value of  sign($|\xi_x|^{\frac{1}{4}},\xi_x)$, where  $\xi_x$ is  the  component of the displacement vector illustrated
	in units of $R_s$.}\label{l3nr0contour} 
\end{figure} 
\begin{figure}*
	\hspace{-1.0cm}	
	\hspace*{-1.4cm}	\includegraphics[angle=0,width=20cm] 
	{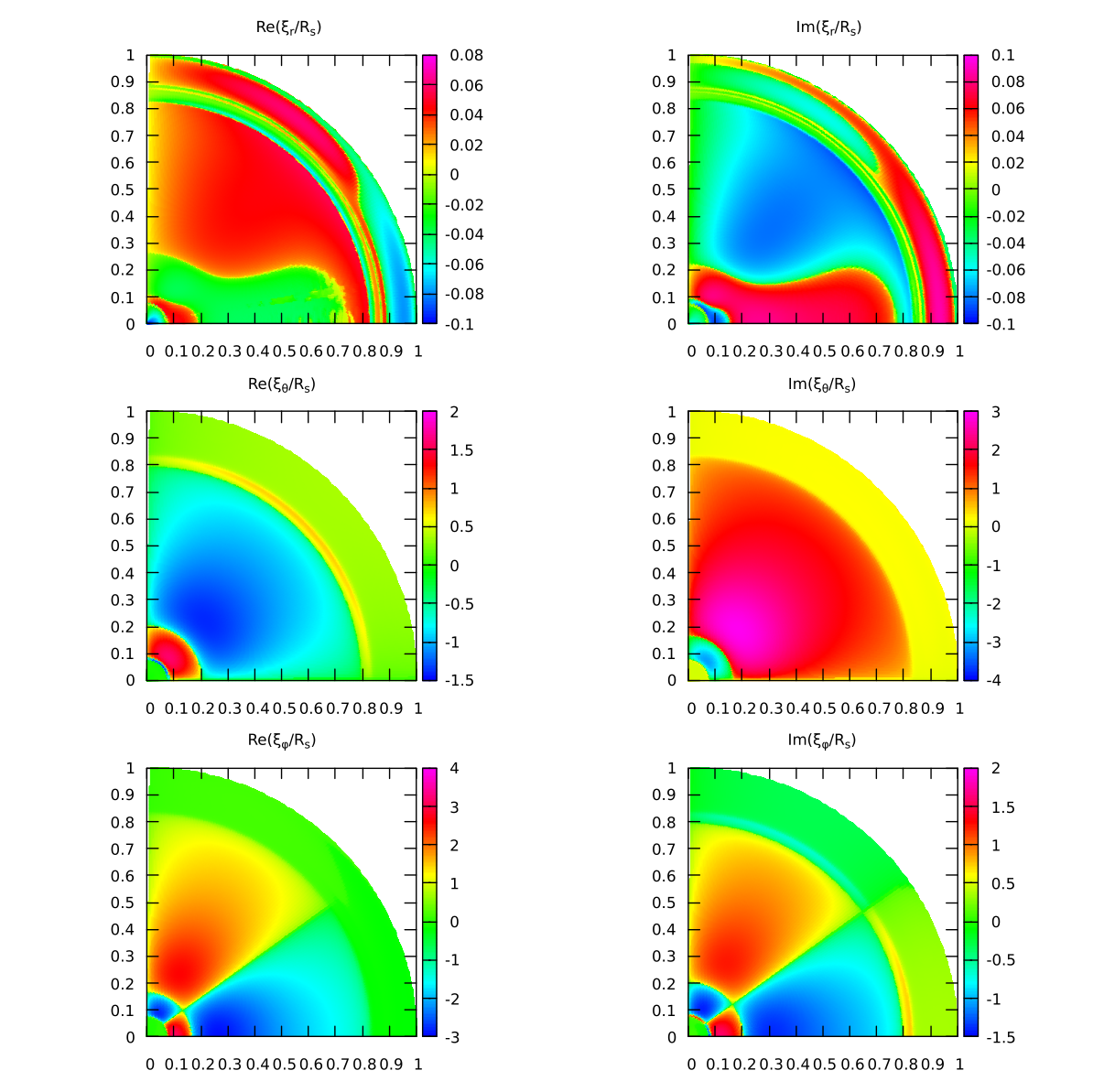}
	\caption{As for Fig. \ref{l3nr0contour} but for the $l'=3$ $r$ mode with $n_r=1$. 
		The   resonance frequency  is $\omega_0=-1.9997204 \times 10^{-3} \,\Omega_c.$
		Compared to the case with $n_r=0,$  the horizontal components of the displacement
		have an extra radial node in the inner radiative region. 
		\label{l3nr1contour} } 
\end{figure} 
	Just as for the $l'=1$ case there is essentially no radial dependence of the horizontal components
	of the Lagrangian displacement  in the convective envelope and 
	the ratio of radial to horizontal displacements is small  characteristically of order $\epsilon_2$ or often smaller at some locations.
	As noted in the discussion below equation (\ref{B1}) in appendix \ref{AS4}, that is expected 
	for free $r$  modes.} 

\textcolor{majenta}{Notably the angular dependence of the horizontal components in the interior is as expected for $r$ modes with $l'=3,$ 
there being no nodes in $\xi_{\theta}$ and one node in $\xi_{\phi}.$ This dependence is seen both in the radiative interior
and the convective envelope.  But note that there is evidence of a spheroidal contribution with $m=2$ and $l'=4$ 
to $\xi_r$ in the radiative interior particularly for the mode with $n_r=0.$ 
The  number of nodes associated with $\xi_{\theta}$ and $\xi_{\phi},$ when viewed as a function of $r,$ in the radiative interior for $n_r=0$ and $n_r=1$
modes follows the same pattern as for  the $l'=1$ modes.}

\textcolor{majenta} {\subsection{ The non resonant  tidal response of the $1.3 M_{\odot}$ star for rotation rates \textcolor{majenta}{ $\Omega_s/\Omega_c$ in the interval  $(5 \times 10^{-3}   , \, 3.5 \times 10^{-2} $) } \label{Nonresresults}
}\label{Genresp}
To determine the tidal evolution of the  spin-orbit misalignment angle, $\beta,$ and the semi-major axis, $a,$ following IP2
we determine the response   produced by forcing  with the tidal potential,  ${\cal U}_{n,m},$ with $(n,m)=(-1,0),$ $(n,m)=(-2,0),$
and $(n,m)=(-2,-2).$
The first two cases with $m=0$ appear stationary in a non rotating frame and only contribute to the tidal evolution
of circular orbits when there is misalignment of spin and  orbital angular momenta.}

\textcolor{majenta}{When  $m=0,$ the forcing frequency is $n\Omega_s.$ This does not depend on $n_0$ and so the response, being linear,  only needs to be calculated
for single  values of  $M_p$ and $a.$ The response amplitude being $\propto  M_p/a^3$ is readily scaled to apply to other values of $M_p$ and $a.$
The range of rotation periods considered  for our model is $(5.88d, 33.15d ).$ This enables us to consider the tidal interactions 
of  a significant number of observed extrasolar planetary systems.
The  value of the centrifugal parameter $\epsilon_2$ is small,  as required  under the assumptions of our modelling.}

\

\begin{table} 
	\begin{center}
		\begin{tabular} {ccccc} 
			\hline 
			$\omega_{f,n,m}/\Omega_c$  &  $E_{kin,n,m}$ (erg) &   Dissp rate (erg/s) & $\Omega_s /\Omega_c$ &  $P_{spin} (d)$ \\ 
			\hline						
-5.500000E-03 & 1.513289E+39 & 6.997529E+21 & 5.500E-03 & 3.31492E+01 \\
-6.500000E-03 & 1.082613E+39 & 5.345359E+21 & 6.500E-03 & 2.80493E+01 \\
-7.500000E-03 & 8.090365E+38 & 3.517746E+21 & 7.500E-03 & 2.43094E+01 \\
-8.500000E-03 & 6.263986E+38 & 2.126141E+21 & 8.500E-03 & 2.14496E+01 \\ 
-1.000000E-02 & 4.487382E+38 & 1.048095E+21 & 1.000E-02 & 1.82320E+01 \\ 
-1.200000E-02 & 3.081027E+38 & 5.178735E+20 & 1.200E-02 & 1.51934E+01 \\ 
-1.400000E-02 & 2.238712E+38 & 3.517302E+20 & 1.400E-02 & 1.30229E+01 \\ 
-1.600000E-02 & 1.695950E+38 & 2.904940E+20 & 1.600E-02 & 1.13950E+01 \\ 
-1.800000E-02 & 1.326548E+38 & 1.650210E+20 & 1.800E-02 & 1.01289E+01 \\ 
-2.000000E-02 & 1.064303E+38 & 1.021029E+20 & 2.000E-02 & 9.11601E+00 \\ 
-2.200000E-02 & 8.717706E+37 & 7.224273E+19 & 2.200E-02 & 8.28729E+00 \\ 
-2.400000E-02 & 7.264647E+37 & 5.600331E+19 & 2.400E-02 & 7.59668E+00 \\
-2.600000E-02 & 6.142357E+37 & 4.542858E+19 & 2.600E-02 & 7.01232E+00 \\ 
-2.800000E-02 & 5.258381E+37 & 3.781305E+19 & 2.800E-02 & 6.51144E+00 \\ 
-3.000000E-02 & 4.550290E+37 & 3.207561E+19 & 3.000E-02 & 6.07734E+00 \\ 
-3.100000E-02 & 4.248108E+37 & 2.971728E+19 & 3.100E-02 & 5.88130E+00 \\ 
			\hline 			
		\end{tabular}
	\end{center}
	\caption{ Results for  forcing of the star with $(n,m)=(-1,0)$: the columns moving from left to right  contain the the dimensionless forcing frequency $\omega_{f,n,m}/\Omega_c$, the kinetic energy,  the  Dissipation rate $-\dot{E}_{kin,n,m} \equiv - \dot{E}_{kin,-n-,m}$,  the dimensionless 
		spin rate $\Omega_s/\Omega_c$ and the spin period in days. }
	\label{tabl11}
\end{table}

\begin{figure}
	\hspace*{-3cm}	\includegraphics[angle=0,width=22cm] 	 
	{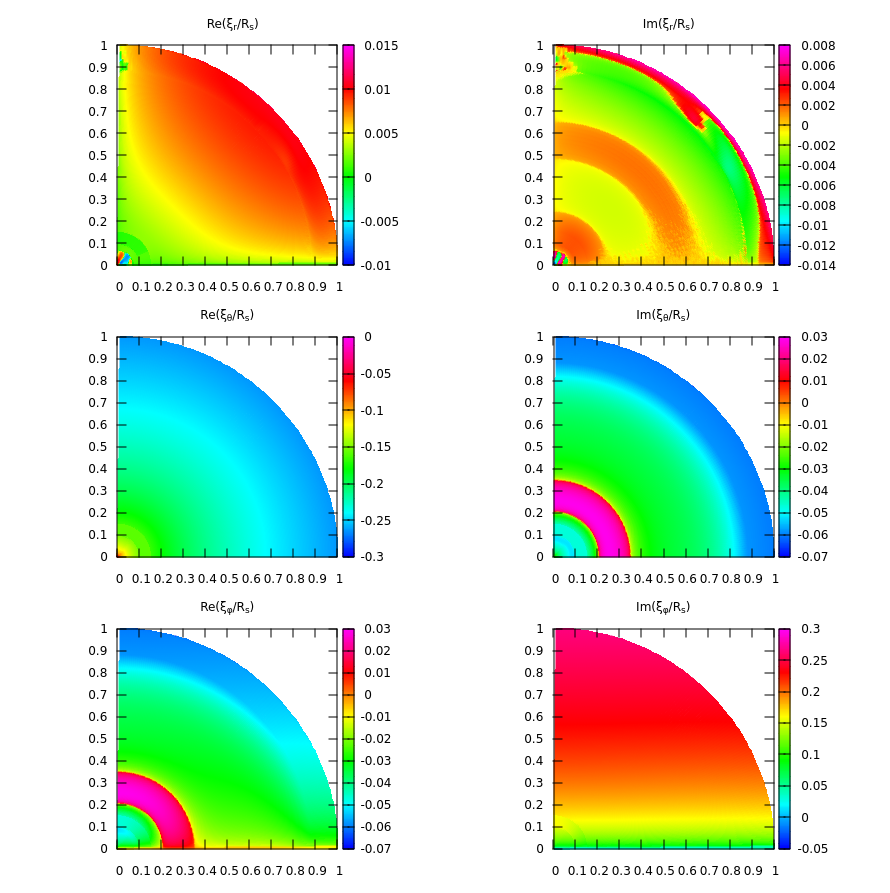}	
	\caption{Contour plots  in the primary's meridional plane $\phi=0$ 
		showing the response to forcing with, ${\cal U}_{-1,0}, (n,m)= (-1,0)),$  for $\Omega_s= 1.00\times 10^{-2} \, \Omega_{c}$ ($P_{spin}=18.23 d$).
		{ For these values of $n$ and $m$ } the forcing frequency is $-\Omega_s$  corresponding to a time independent perturbation in the non rotating frame. 
		{This calculation}  was done for $a/R_s=10.$ The response for other values can be obtained by applying the scaling factor $(a/R_s)^3.$ 
		The Cartesian coordinates along the two axes indicate the relative radius $r/R_s$. 
		The vertical colour bars on the right indicate the local value of  sign($|\xi_x|^{\frac{1}{4}},\xi_x)$, where  $\xi_x$ is  the  component of the displacement vector illustrated.  \label{n-1m0contour}}
\end{figure}

\begin{figure}
	\hspace{0.2cm}	\includegraphics[height=10cm, width=12cm]
	{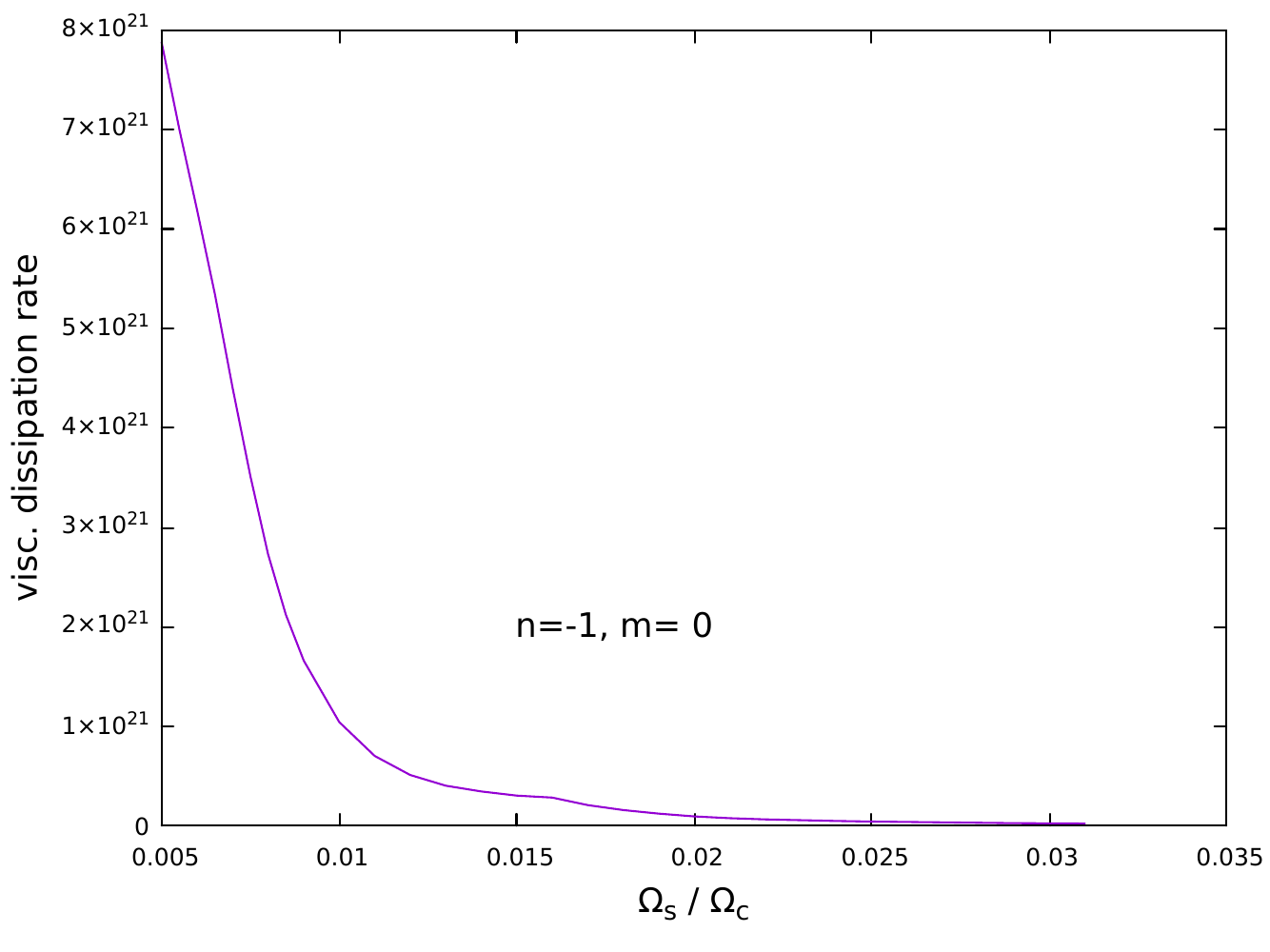}
	
	\hspace{0.25cm}	\includegraphics[height=10cm, width=12cm]
	{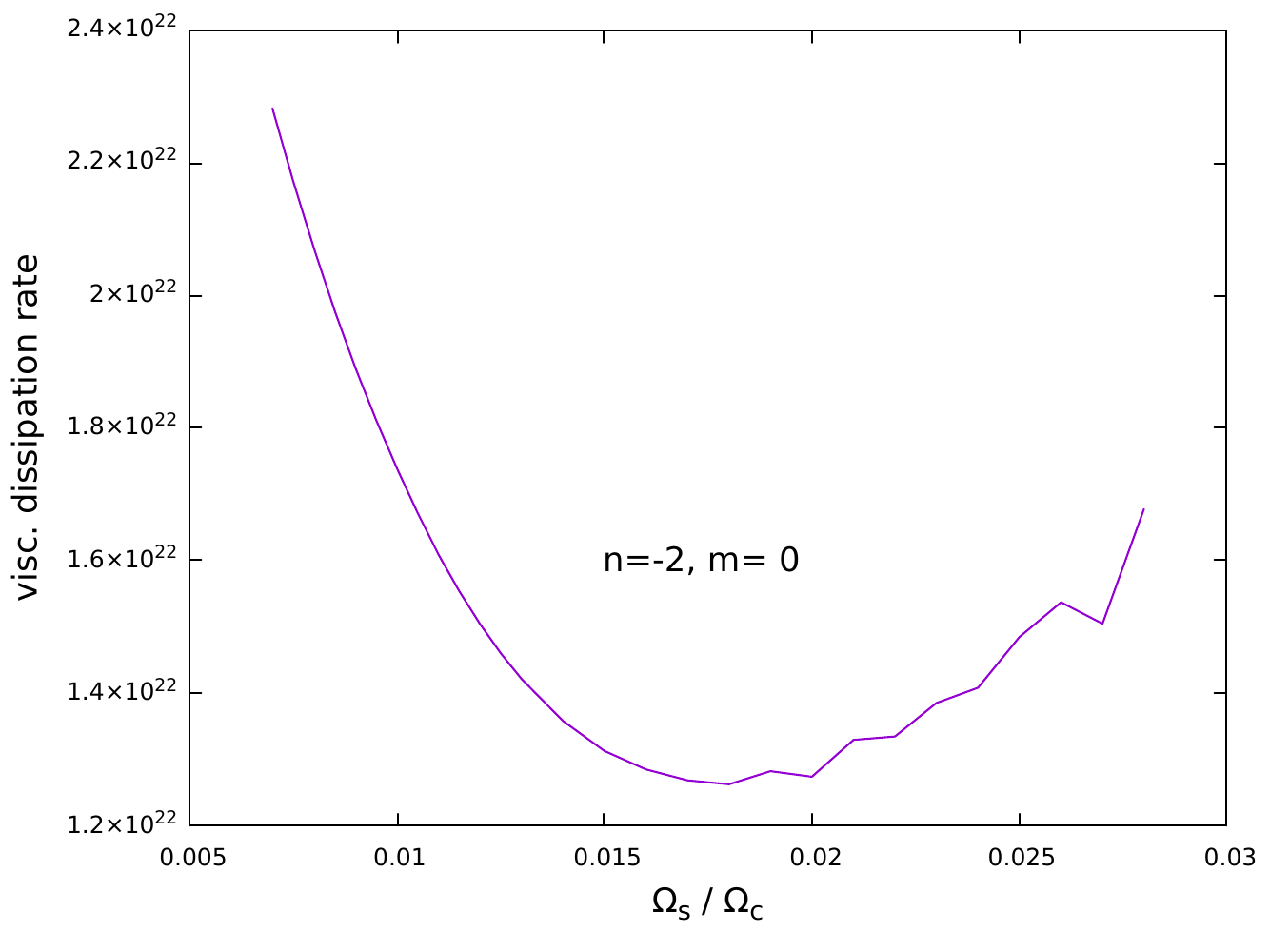}
	\caption{ \color{majenta} The upper panel shows the viscous dissipation rate in the convective envelope, 	 $ -\dot{E}_{kin,-1,0} \equiv -\dot{E}_{kin,1,0}$ (erg/s),  for $ (n,m)= (-1,0)$  forcing  as a function of the stellar spin rate.  
		The lower panel shows the same but  for forcing with $(n,m)=(-2,0)$. In each case the the forcing was by a Jupiter mass planet with $a/R_s =10.$
		Note that these dissipation rates are $\propto (a/R_s)^6$  which can be used for scaling. }
	\label{Dispraten-1n-2m0}
\end{figure}

\begin{table} 
	\begin{center}
		\begin{tabular} {ccccc} 
			\hline
			$\omega_{f,n,m}/\Omega_c$  &  $E_{kin,n,m}$ (erg) &   Dissp rate (erg/s) & $\Omega_s /\Omega_c$ & $P_{spin} (d)$ \\ 			 
			\hline
-1.400000E-02 & 4.969208E+29 & 2.281638E+22 & 7.000E-03 & 2.60458E+01 \\
-1.600000E-02 & 4.754501E+29 & 2.071177E+22 & 8.000E-03 & 2.27900E+01 \\
-1.800000E-02 & 4.685754E+29 & 1.891274E+22 & 9.000E-03 & 2.02578E+01 \\
-2.000000E-02 & 4.729188E+29 & 1.739436E+22 & 1.000E-02 & 1.82320E+01 \\
-2.200000E-02 & 4.867349E+29 & 1.609986E+22 & 1.100E-02 & 1.65746E+01 \\
-2.400000E-02 & 5.092210E+29 & 1.505268E+22 & 1.200E-02 & 1.51934E+01 \\
-2.600000E-02 & 5.399610E+29 & 1.422379E+22 & 1.300E-02 & 1.40246E+01 \\
-3.000000E-02 & 6.262151E+29 & 1.313705E+22 & 1.500E-02 & 1.21547E+01 \\
-3.400000E-02 & 7.477276E+29 & 1.269540E+22 & 1.700E-02 & 1.07247E+01 \\
-3.800000E-02 & 9.066072E+29 & 1.283183E+22 & 1.900E-02 & 9.59581E+01 \\ 
-4.200000E-02 & 1.074250E+30 & 1.330370E+22 & 2.100E-02 & 8.68192E+00 \\ 
-4.600000E-02 & 1.233267E+30 & 1.386050E+22 & 2.300E-02 & 7.92697E+00 \\ 
-5.000000E-02 & 1.424094E+30 & 1.485526E+22 & 2.500E-02 & 7.29281E+00 \\ 
-5.400000E-02 & 1.641599E+30 & 1.505591E+22 & 2.700E-02 & 6.75260E+00 \\ 
-5.600000E-02 & 1.755753E+30 & 1.678174E+22 & 2.800E-02 & 6.51144E+00 \\ 
			\hline 			
		\end{tabular}
	\end{center}
	\caption{ Results for  forcing of the star with $(n,m)=(-2,0)$: the columns moving from left to right  contain the the dimensionless forcing frequency $\omega_{f,n,m}/\Omega_c$, the kinetic energy,  the  Dissipation rate $-\dot{E}_{kin,n,m} \equiv - \dot{E}_{kin,-n-,m}$,  the dimensionless 
		spin rate $\Omega_s/\Omega_c$ and the spin period in days. }
	\label{n-2m0contour}	\label{tabl12}
\end{table}
\textcolor{majenta}{For $(n,m)$=(-1,0)  the forcing frequency is $-\Omega_s.$  This is close to the  $r$ mode  resonances with $l'=1$ with $n_r=0,$ and $n_r=1.$
However, we recall that the half-widths of these resonances $\sim 10^{-10} \,\Omega_c$ is much smaller than the separation of the mode frequencies
from $\Omega_s,$ the latter being $\sim 1000$ times larger. As indicated above,  this   results  in the response with $n=-1$ and $m=0$ being effectively  non resonant as far as these modes
are concerned.
}
We determined the  response  of  the $1.3M_{\odot}$ 
star  to  a Jupiter mass planet in a circular orbit  with,
$a/R_s=10,$
the  orbital period then being $5.763d.$  For this value of $a/R_s,$ values of  \textcolor{majenta}{ $\Omega_s/\Omega_c$ in the interval  $(5.5 \times 10^{-3}   , \, 3.1 \times 10^{-2} $) }  were considered.
\textcolor{majenta}{
\subsubsection{Response to forcing with $(n,m)=(-1,0).$} }
Contour plots of the Lagrangian displacement in the stellar mid-plane for $ (n,m)= (-1,0)$ at the representative angular velocity 
$\Omega_s= 1.00\times10^{-2} \,\Omega_{c}$ ($P_{spin}=18.23 \,d$)
are shown in in Fig. \ref{n-1m0contour}.
 Although far in the wings of the resonances,
 these contour plots have global  features similar to those seen in Figs. \ref{l1nr0contour} and \ref{l1nr1contour}.
In particular the horizontal components of the Lagrangian displacement show minimal radial dependence in the convective envelope. In addition  ratio of  the magnitudes 
of  the radial and  horizontal displacements, is similar to that  for the free modes,
being  typically ${\rm O}(\epsilon_2).$ The viscous dissipation rate in the convective envelope, 
$ -\dot{E}_{kin,-1,0} \equiv -\dot{E}_{kin,1,0},$   is shown  as a function of the stellar spin rate in the upper panel of Fig. \ref{Dispraten-1n-2m0}.
The results are also presented in tabular form in table \ref{tabl11}.
The characteristic magnitude of this quantity, being $10^{21-22},$ is very much less than the same quantity at the  $(l'=1, n_r =0)$ resonance peak $\sim 10^{27},$
thus being located far into the resonance wings. However, the  increase of  $-\dot{E}_{kin,-1,0}$  with decreasing $\Omega_s$ is consistent with the relative separation
from resonance decreasing with $\Omega_s$ (see  discussion in Section \ref{l1nr0resonanceshift}). 

\subsubsection{Response to forcing with $(n,m)=(-2,0)$}\label{2,0response}
\begin{figure}
	\hspace*{-3cm}	\includegraphics[angle=0,width=22cm] 	
	{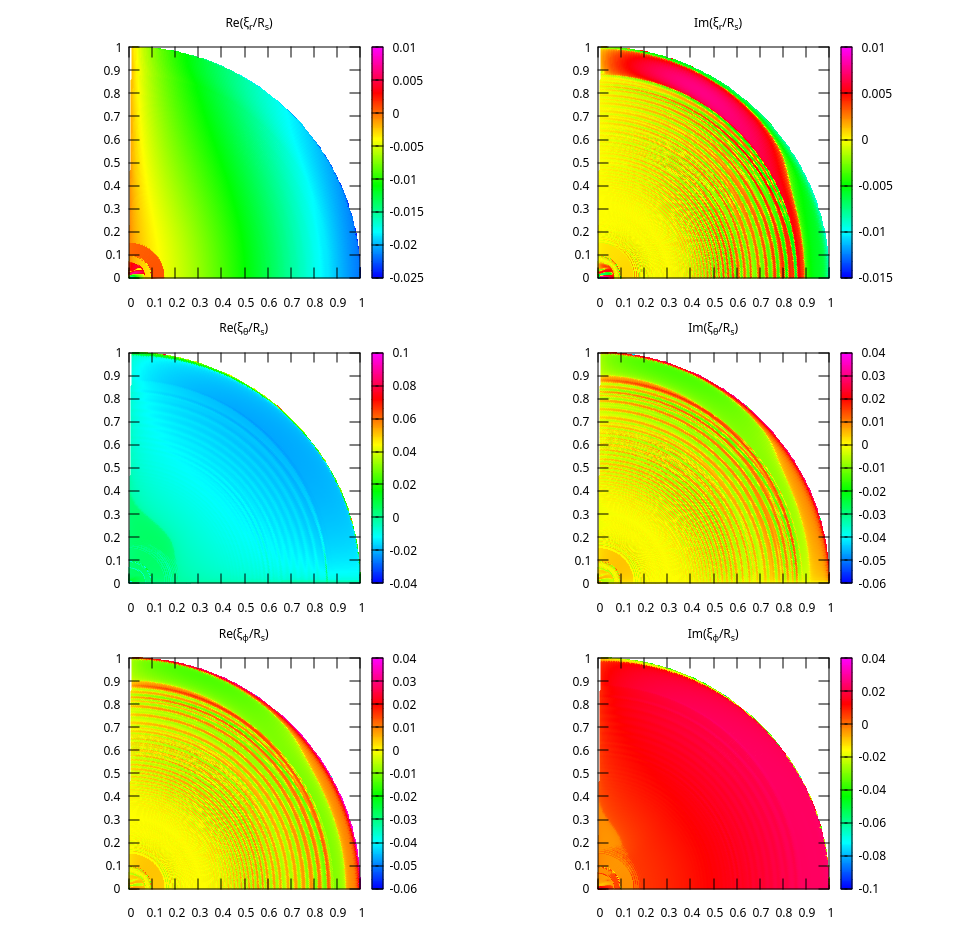}		
	\caption{As in Fig. \ref{n-2m0contour} but showing contour plots  in the primary's meridional plane $\phi=0$ 
		illustrating the response to forcing with, ${\cal U}_{-2,0}, (n,m)= (-2,0)),$  for 
		$\Omega_s=2.73\times 10^{-2} \,\Omega_c$ ($P_{spin} = 6.678d$).}
		\label{n-2m0contour2}
\end{figure}

\textcolor{majenta}{ We also evaluated the response of the  $1.3 M_{\odot}$ star  to forcing  with the tidal potential ${\cal U}_{n,m}$
for  $(n,m) =(-2,0)$ for the same interval in $\Omega_s.$ Contour plots illustrating the components of the Lagrangian displacement for the case with 
$\Omega_s=1.00\times 10^{-2} \,\Omega_c$ ($P_{spin} = 18.23d$)  are presented in Fig. \ref{n-2m0contour} 
\begin{figure}
	\hspace*{-3cm}	\includegraphics[angle=0,width=22cm] 		
	{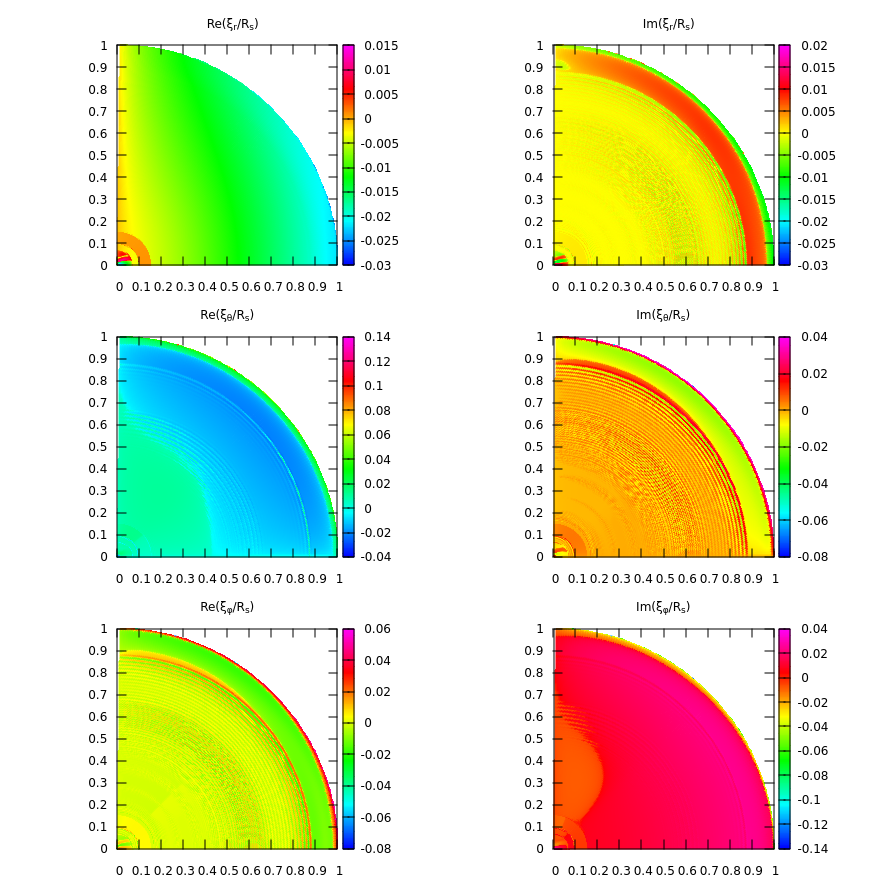}	
	\caption{As in Fig. \ref{n-1m0contour} but showing contour plots  in the primary's meridional plane $\phi=0$ illustrating the response to forcing with ${\cal U}_{-2,0}, (n,m)= (-2,0)$  for 
		$\Omega_s=1.00\times 10^{-2} \,\Omega_c$ ($P_{spin} = 18.23 d$).}
	\label{n-2m0contour}
\end{figure} 
and for comparison the corresponding plot for  more rapid rotation with
$\Omega_s=2.73\times 10^{-2}\,\Omega_c.$ ($P_{spin} = 6.678d$)
is given  in  Fig. \ref{n-2m0contour2}.}
\textcolor{majenta}{In this case the forcing frequency of $-2\Omega_s$ is even more distant from an $r$ mode resonance than the
previous case with $n=-1,$ and $m=0.$ A consequence is that the angular dependence of the horizontal components of the Lagrangian displacement
appears mainly spheroidal. The typical relative  magnitude of the radial component of the displacement as compared to that of the horizontal component
is significantly larger \textcolor{majenta}{ than for a $r$ mode}. This is as expected from the discussion in Section \ref{epicyclic} which indicates that they can be of comparable magnitude,
The presence of relatively large radial displacements near the inner boundary of the convective envelope suggests the possibility of the excitation
of rotationally modified $g$ modes that propagate into the inner radiative region.
Short wave length perturbations associated with such  modes are visible in Figs \ref{n-2m0contour} and \ref{n-2m0contour2}, being more prominent
in the latter case with shorter rotation period.
The implementation of the artificial viscosity, $\nu_{min}= 10^{9}$ in cgs units, was necessary in order to maintain numerical stability in these runs. }

\textcolor{majenta}{ The viscous dissipation rate in the convective envelope, 
$ -\dot{E}_{kin,-2,0} \equiv -\dot{E}_{kin,2,0},$   is shown   as a function of the stellar spin rate in the lower panel of Fig. \ref{Dispraten-1n-2m0}.
The results are also presented in tabular form in table \ref{tabl12}.
It is notable that $ -\dot{E}_{kin,-2,0} $ exceeds the corresponding dissipation rate for $(n,m)=(-1,0)$
for all values of  $\Omega_s/\Omega_c $ considered. 
This is particularly marked for $\Omega_s/\Omega_c >\sim 0.02.$ }

\subsubsection{Response to forcing with $(n,m)=(-2,-2)$}\label{-2,-2}
\textcolor{majenta}{Finally we evaluated the response of the  $1.3 M_{\odot}$  star  to forcing  with the tidal potential ${\cal U}_{n,m}$ for  $(n,m) =(-2,-2)$ within the same interval of $\Omega_s.$  In this case the forcing frequency is $2(n_0-\Omega_s)$ with  the orbital period  $5.76d$ corresponding to $a/R_s=10.$ It is forcing  with these values of $m$ and $n$
that is mainly responsible for driving a circular orbit towards synchronisation and it survives in the non rotating limit.
Contour plots illustrating the components of the Lagrangian displacement for the case with 
$\Omega_s=8.36\times 10^{-3}\, \Omega_c$ ($P_{spin}=21.81d$)  are presented in in  Fig. \ref{n-2m-2contour}.  
\begin{figure}
	\hspace{-3.0cm}	
	\includegraphics[angle=0,width=1.2\columnwidth]  {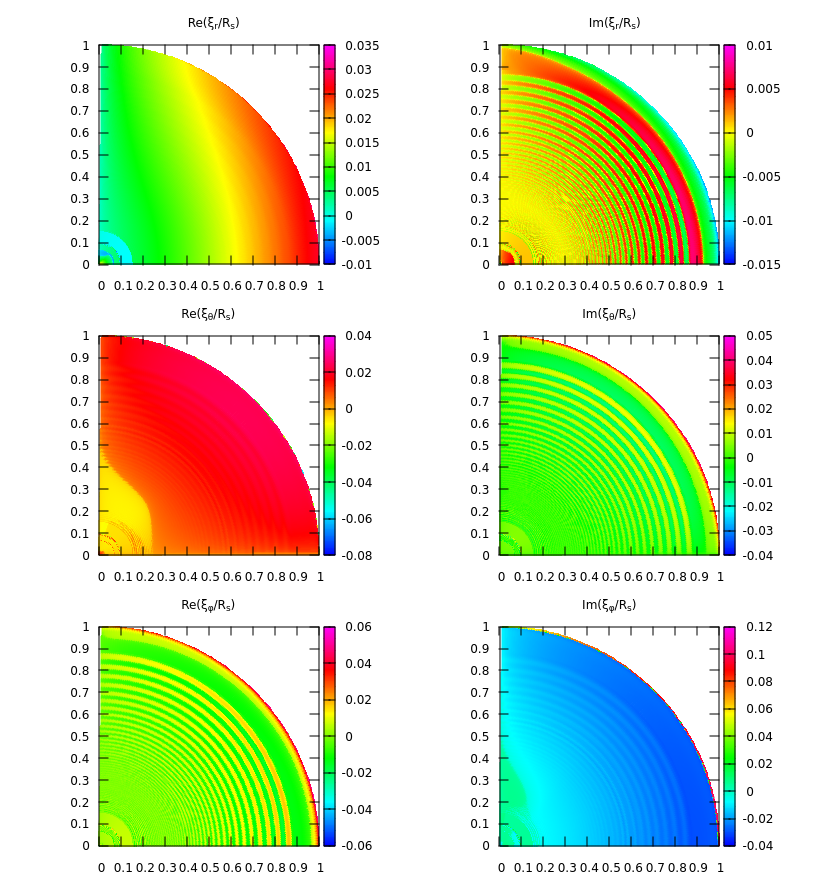} 
	\caption{ As in Fig. \ref{n-1m0contour} but showing contour plots  in the primary's meridional plane $\phi=0$ 
		illustrating the response to forcing with, ${\cal U}_{-2,-2}, (n,m)= (-2,-2)),$  for 
		 $\Omega_s=8.36\times 10^{-3}\Omega_c$ ($P_{rot}=21.81d$).
		 The orbital period was 5.76d corresponding to $a/R_*=10.$ }
		 \label{n-2m-2contour}
\end{figure} 
In this case the angular dependence of the forcing is the same as that for  $n=-2,m=0. $
\footnote{Though both the forcing frequency and the forcing potential have the opposite sign} 
Accordingly the form of the angular dependencies seen in the contour plots is similar. However, the rotationally modified $g$ modes in the radiative interior
are  prominent as  expected for the relatively high forcing frequency in this case, which is similar to that for the case illustrated in Fig. \ref{n-2m0contour2}.
The viscous dissipation rate in the convective envelope $ -\dot{E}_{kin,-2,-2} \equiv -\dot{E}_{kin,2,2}$ is shown   as a function of the stellar spin rate in  Fig.\ref{Disprate2an-2m-2}.   In this case the form of the dissipation rate as a function of  $\Omega_s$ is similar to that found when ${n,m}= (-2,0),$ while being of 
 of comparable magnitude.}
 
\textcolor{majenta}  {It is expected to tend to a limit as $\Omega_s\rightarrow 0$ corresponding to a non rotating star. 
\begin{figure}
	\includegraphics[height=10cm, width=12cm]
	{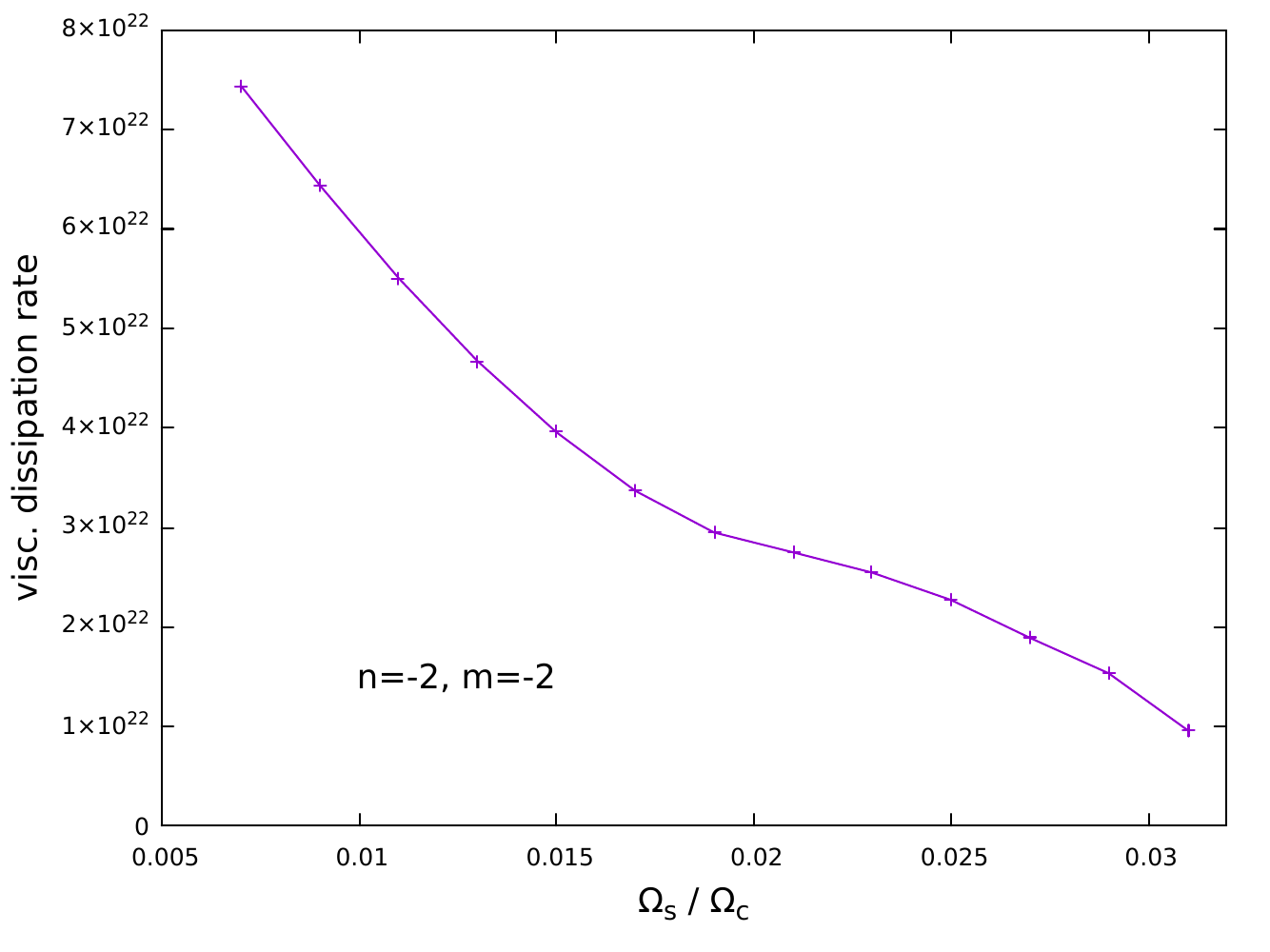}	
	\caption{The viscous dissipation rate $ -\dot{E}_{kin,-2,-2} \equiv -\dot{E}_{kin,2,2}$ (erg/s) in the convective envelope for $(n,m)= (-2,-2)$
		\textcolor{majenta}{	forcing  by a Jupiter mass planet with $P_{orb}= 5.76d$  as a function of the stellar spin rate for $1.3M_{\odot}.$}} \label{Disprate2an-2m-2}
\end{figure}
For $\Omega_s > 1.582 \times 10^{-2} \Omega_c$, corresponding to $P_{spin} < 11.5 \, d$, the tidal forcing enters the inertial range. For small $\Omega_s$ the dissipation rate is 
expected to vary linearly with $\Omega_s$ or $1/P_{spin}$. 
The spin and orbit are synchronised when the spin period is $5.76d$ and we would expect the energy dissipation rate to approach zero as this state is approached, though numerical issues associated with short wavelength responses prevent approaching this too closely.}
{
Noting the above two considerations, we find that the simple linear expression
\begin{align}
-\dot{E}_{kin,-2,-2} = -3.779 \times 10^{24} (\Omega_s/\Omega_c) + 9.821 \times 10^{22}
\end{align}
fits the solution shown in Fig. \ref{Disprate2an-2m-2} for $\Omega_s/\Omega_c < 0.019 $  (corresponding to spin periods  larger than 9.6  d )
with a relative error $ < \sim 10\%.$}
 
\noindent   

\begin{figure}
	\hspace*{-3cm}	\includegraphics[angle=0,width=22cm] 		
	{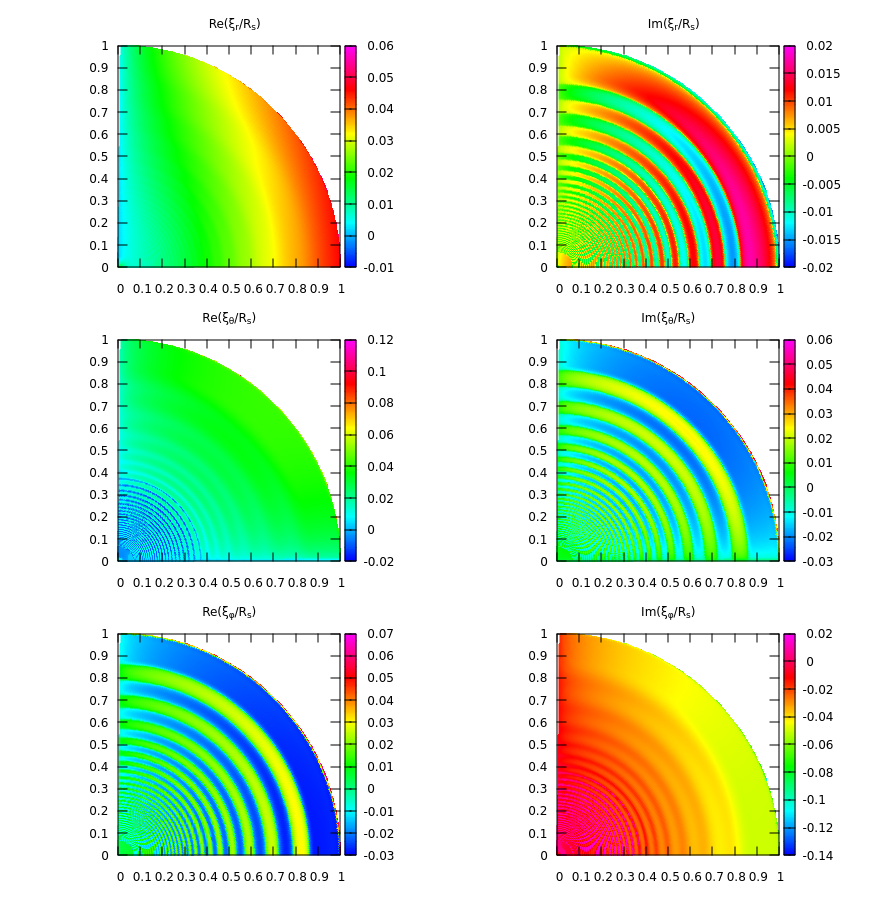}		
		\caption{ As in Fig. \ref{n-2m0contour2} but now for tidal forcing with $a/R_s)=5$ corresponding to a much smaller orbital period of $2.038 d$ with $\Omega_s=1.236\times 10^{-2} \,\Omega_c$ ($P_{spin}=14.75d$) with a prominent lower order rotationally modified $g$ mode response in the core due to a much higher forcing frequency $\omega_{f,-2,-2}=0.1542311 \,\Omega_c$.}
		\label{n-2m-2contour3}
	\end{figure}

It should be noted that, contrary to the $m=0$ forcing, the tidal dissipation rate cannot be simply scaled with $a^{-6}$ for other orbital periods since the forcing frequency now depends on $n_o$. However, it appears that  for orbital periods between $10.6 \, d$ ($a/R_s=15)$ and $2.04\, d$ $(a/R_s=5)$ the calculated dissipation rates shown in Fig. \ref{Disprate2an-2m-2} when scaled with $a^{-6}$ are adequate for order of magnitude estimates. The 
  deviation  at the smallest values of $\Omega_s/\Omega_c$  after scaling the results in Fig.\ref{Disprate2an-2m-2}, is a  factor 7-9 at short orbital periods $( \sim 2 \, d)$ for which the higher forcing frequency excites lower radial order rotationally modified $g$ modes in the core, see Fig.\ref{n-2m-2contour3}. This leads to stronger orbital decay of the system.


\clearpage

\section{Effects on tidal evolution}\label{Effecttidalevol}
We make use of  equation (34)  of PS1 for the rate of change of the semi-major axis of a circular orbit in the form
\begin{align}
\left\langle\frac{da}{dt}\right\rangle = -\frac{16  n_0 a^2}{GM_*M_p}  
\sum^{j=2}_{j=-2}\frac{\left(d^{(2)}_{j,2}\right)^2(dE_{kin, j,2}/dt)}{\omega_{f,j,2}} .
\label{adotaveedot}
\end{align}
Here $d^{2}_{j,k}(\beta), {\rm for}\  j = -2, -1, 0, 1, 2,\  {\rm and}\  k=-2, -1, 0, 1, 2$ give the elements of the Wigner small $d$ matrix. These are given in an  appendix in  PS1.\\
In addition we  recall that $\omega_{f,j,m}=-mn_o+j\Omega_s.$

In the limit $\Omega_s\rightarrow 0$ the  quantities, $(dE_{kin, j,2}/dt)/{\omega_{f,j,2}}$ are independent of $j$ on account of the spherical symmetry
of the stellar model in this limit (see PS1). In addition $\displaystyle{ \sum^{j=2}_{j=-2}\left(d^{(2)}_{j,2}\right)^2=1 } .$ Thus in this limit
equation (\ref{adotaveedot}) becomes
\begin{align}
\frac{1}{a}\left\langle\frac{da}{dt}\right\rangle = \frac{8  a}{GM_*M_p}  
\left(\frac{ dE_{kin, 2,2}}{dt}\right) .
\label{adotaveedot1}
\end{align}
This leads to
\begin{align}
\frac{a}{\left\langle{da}/{dt}\right\rangle} = 1.66\times10^{13} 
\left(\frac{14.8R_{\odot}}{a}\right)\left(\frac{M_*}{1.3M_{\odot}}\right)\left(\frac{M_p}{10^{-3}M_{\odot}}\right)  
\left(\frac{8.0\times 10^{22}cgs}{ dE_{kin, 2,2}/dt}\right)y ,
\label{adotaveedot1a}
\end{align}
where parameters have  been scaled to those for a Jupiter mass planet with $a/R_s=10.$ 
Note that from the discussion in Section \ref{-2,-2}, as well as the consideration of tides in general,
we expect this time scale to increase as the stellar spin is increased towards synchronisation.
We note that equation (\ref{adotaveedot1a}) leads to a time scale significantly longer by almost two orders of magnitude
than that obtained from equation (6.2) of \citet{Zahn1977}
when adjusted to apply to synchronisation rather than circularisation. This is probably related to the small convective envelope in this case.

\subsection{Evolution of the spin orbit angle}\label{spinorbevol}
Furthermore we make use of equation (38) of PS1 for the mean rate of change
of the angle between the orbital and spin angular momenta, $\beta$ in the form
\begin{align}
\left\langle \frac{d \beta }{dt}\right\rangle  =&
-\frac{4  n_o  a \left(1+L\cos\beta/S\right)
}{G M_pM_*}
\sum^{j=2}_{j=-2} \left(\frac{d^{(2)}_{j,1}d^{(2)}_{j,2} (dE_{kin, j,2}/dt)}{\omega_{f,j,2}}
+\sqrt{\frac{3}{2}}\frac{d^{(2)}_{j,1}d^{(2)}_{j,0}(dE_{kin, j,0}/dt)}{\omega_{f,j,0}}
\right)\nonumber\\
&-\frac{8\sin\beta}{S} 
\sum^{j=2}_{j=-2}\frac{\left(d^{(2)}_{j,2}\right)^2(dE_{kin, j,2}/dt)}{\omega_{f,j,2}}
\label{Gaussequation4.3y}
\end{align}

We note that the 
$\beta$ dependence in equations (\ref{adotaveedot}) and (\ref{Gaussequation4.3y}) is contained entirely within the
Wigner small $d$  matrix elements. In the  case of, $a,$  the rate of evolution is determined by the rate of kinetic energy changes,  $dE_{kin.,j,2}/dt,$ 
for $j=-2,-1,0,1,2.$ In the case of the inclination, the evolution  depends on these together with values of,  $ dE_{kin, j,0}/dt,$ for $j=1$ and $j=2,$ 
noting that for $m=0,$  contributions from  terms with $j < 0$
can be found from those  with  $j > 0,$ using  the relation (\ref{dotER}). 
There is no contribution with   $m=j=0.$   

We remark that,  by making use  of  equation (\ref{adotaveedot}),  it becomes apparent that
 the term incorporating  the second summation, being the last term  on the right hand side of (\ref{Gaussequation4.3y}), can be expressed as
\begin{align}
\frac{\sin\beta}{S}\left\langle\frac{dL}{dt} \right\rangle.\label{Ldott}
\end{align}
It accordingly represents the mean rate of change of the spin orbit angle, $\beta,$ resulting from angular momentum exchange
between  orbit and spin occurring  through  tidal interaction,   that is necessitated by total  angular momentum  conservation. The first term in the first summation
on the right hand side of (\ref{Gaussequation4.3y}) involves  a summation of contributions with $m=2.$
In PS1 we argued that these amount to  a negligible contribution when $\Omega_s/n_0$ is small. However, this situation may not apply to stars near the Kraft
break that are expected to rotate more rapidly than a solar type star as considered in PS1. However, we can find a bound 
$\propto  \left\langle{dL}/{dt} \right\rangle$ for these contributions applicable when there is significant misalignment.


\subsection{A useful bound when there is significant misalignment}\label{bound}
Consider the ratio
\begin{align}
&{\cal X}\quad =\quad\frac{\displaystyle \sum^{j=2}_{j=-2} \frac{d^{(2)}_{j,1}d^{(2)}_{j,2} (dE_{kin, j,2}/dt)}{\omega_{f,j,2}}}
{\displaystyle \sum^{j=2}_{j=-2}\frac{\left(d^{(2)}_{j,2}\right)^2(dE_{kin, j,2}/dt)}{\omega_{f,j,2}}} \quad
= \quad \frac{\displaystyle \sum^{j=2}_{j=-2}\alpha_j x_j }
{\displaystyle \sum^{j=2}_{j=-2} x_j
}
\nonumber\\
\label{sinequality}
\end{align}
where $x_j = \left(d^{(2)}_{j,2}\right)^2(dE_{kin, j,2}/dt)/{\omega_{f,j,2}},$ and 
$\alpha_j =  d^{(2)}_{j,1}/{d^{(2)}_{j,2}}.$ 
It can be verified that  the ratio of the magnitude of the contribution of the  first term in the summation  that occurs in the first expression on the right hand side of (\ref{Gaussequation4.3y})
to  the last term, the latter as indicated above,  being $\propto  \left\langle{dL}/{dt} \right\rangle,$ is
\begin{align}
&{\cal X}_R= \frac{{\cal X}}{2\sin\beta}\left(\frac{S}{L}+\cos\beta\right).
\label{sratio}
\end{align}

We shall assume that $\omega_{f,j,2} <  0$ and thus the system is sub-synchronous.
Then given that $dE_{kin, j,2}/dt <  0, $we  have $x_j >0.$
It then follows that
\begin{align}
&|{\cal X}|\quad \le \quad \displaystyle \sum^{j=2}_{j=-2}|\alpha_j| x_j 
\bigg{/}
\left(\displaystyle \sum^{j=2}_{j=-2} x_j\right) <   \sqrt{\displaystyle \sum^{j=2}_{j=-2}\alpha_j^2}
\end {align}

Making use of  the small $d$ Wigner matrix elements given in PS1 we find that
for $\beta=\pi/4$ or $\beta = 3\pi/4,$ we find $|{\cal X}|< 2\sqrt{10}$ and for $\beta=\pi/2$ we obtain  $|{\cal X}|< \sqrt{10}.$ 
From (\ref{Ldott}) and  (\ref{sratio}) the ratio of the contribution arising first term discussed above to the last term
does not exceed $\sqrt{10}(\sqrt{2}S/L+1),$ and $\sqrt{2.5}S/L$ for each of these cases. 

Given the lack of orbital decay observed in hot Jupiter systems, apart from possibly those with short orbital periods
and  the estimated  timescale given by (\ref{adotaveedot1a})
and in addition  that $S/L$ is generally significantly less than unity (see below) we may conclude that when the mean rate 
of change of the  orbital angular momentum can be neglected, the only terms  that could affect
alignment are those involving $m=0$ in (\ref{Gaussequation4.3y}) which correspond to secular forcing. 
These have  often been relied upon for causing alignment for solar type stars and discussed extensively
in PS1. We now consider their operation for hotter stars of the type considered here that are close to the Kraft break. 
In doing this we neglect any stellar spin down that might result from stellar winds \citep[eg.][]{Skum72} as this process is not expected to
be efficient beyond the Kraft break.

\subsection{Evolution of the spin orbit angle induced by secular forcing terms with $m=0$} \label{m0evol}
The mean rate of change of $\beta$  arising from secular forcing terms with $m=0$ follows as 
their contribution to equation (\ref{Gaussequation4.3y}).    Making use of the components of the  small $d$ matrix given in PS1, 
as shown in PS1
this can be reduced to equation (41) of SP1 which reads
\begin{align}
\left\langle \frac{d \beta }{dt}\right\rangle  =&
\frac{6  n_o  a \left(1+L\cos\beta/S\right)\sin\beta
}{G M_pM_*\Omega_s}\left( \cos^2\beta
\frac{dE_{kin, 1,0}}{dt}+\frac{1}{4}\sin^2 \beta \frac{dE_{kin, 2,0}}{dt}\right)
\label{Gaussequation4.3z2}
\end{align}

Now may write 
\begin{align}
\frac{L}{S}= \frac{M_pM_*}{M_p+M_*}\frac{\sqrt{G(M_p+M_*)a}}{I\Omega_s}\approx \frac{M_p}{kM_*} \frac{n_0}{\Omega_s}\left(\frac{a}{R_s}\right)^2=
2.5\left( \frac{40M_p}{kM_*}\right) \frac{n_0}{\Omega_s}\left(\frac{a}{10R_s}\right)^2.\label{LoverS}
\end{align}
We see from (\ref{LoverS})  that being $\propto 1/\Omega_s,$  $L/S$ 
is significantly smaller for the stellar model considered here  as compared to solar type stars on account of their expected more rapid rotation
(see table \ref{table0}).
From (\ref{Gaussequation4.3z2}) we find
\begin{align}
\left\langle \frac{d \beta }{dt}\right\rangle  =&
\frac{( 5.89 \times10^{13}y)^{-1}(10^{3} M_p/M_{\odot})
\left(1+L\cos\beta/S\right)\sin\beta
}{(M_*/(1.3M_{\odot}))}
\left(\frac{10R_s}{a}\right)^{5}\left( \frac{n_0}{\Omega_s}\right)\nonumber\\
&\left( \cos^2\beta
\left( \frac{dE_{kin, 1,0}/dt}{3\times 10^{22}cgs}\right)+\frac{1}{4}\sin^2 \beta \left(\frac{dE_{kin, 2,0}/dt}{3\times 10^{22}cgs}\right)\right).
\label{Gaussequation4.3z2a}
\end{align}
Here  $dE_{kin, 1,0}/dt$  and $dE_{kin, 2,0}/dt$ are illustrated as functions of $\Omega_s/\Omega_c$ in the interval  $(0.0055 , 0.031)$
in the upper and lower panels of Fig. \ref{Dispraten-1n-2m0} respectively,
with numerical values being  given in tables  \ref{tabl11} and \ref{tabl12} respectively.
Note too that we have taken into account that the right hand sides of equations (\ref{Gaussequation4.3z2}) and  (\ref{Gaussequation4.3z2a})
can be scaled to be applicable to different $M_p$ and $a$ through proportionality
to $M_p^2/a^6.$ This enables us to consider orbital periods shorter than $5.76d$ for a prescribed value of $\Omega_s.$

Note that the fact that  absolute magnitude of  $dE_{kin, 2,0}/dt$ significantly  exceeds that of $dE_{kin, 1,0}/dt$. This together with the dependence
on $\beta$ in equation (\ref{Gaussequation4.3z2a}) implies that the relative decay rate of $\beta$ can be significantly more rapid for non negligible values
$\sim \pi/4$ than very small ones.
 While noting the results shown 
 in Figs. \ref{Dispraten-1n-2m0} and \ref{Disprate2an-2m-2} together with tables  \ref{tabl11} and \ref{tabl12},
  we estimate the characteristic time  for $\beta$  to change when there is significant misalignment to be 
$t_{\beta}~=~|(\left\langle {d \beta }/{dt}\right\rangle)|^{-1}.$ 
For a Jupiter mass planet with $a/R_s=10$ corresponding to 
an orbital period of $5.76d$ around a $1.3M_{\odot}$ star we find from 
 (\ref{Gaussequation4.3z2a}) that, 
 \begin{align}
t_{\beta}  >\sim 2.4\times 10^{14}/(\sin^3\beta((1+2.5n_0\cos\beta/\Omega_s)n_0/\Omega_s) y\label{tblimit}
\end{align} 
 for the rotation periods considered.
Thus negligible evolution is expected in this case unless the rotation rate is unrealistically extremely  small.
\\

\subsection{Systems with orbital periods in the range $2.8-5$ days}\label{2.8days}
We recall that  as discussed above, through an appropriate scaling,  (\ref{Gaussequation4.3z2a})  is applicable for arbitrary $a.$
We make use of this feature in the discussion  below.
Table \ref{table0} indicates that WASP 7 has an orbital period of $4.95 d$  and  \citet{Serrano} 
find a stellar rotation period of $5.18d.$
In addition  observations of  $V\sin i$   indicate a rotation period in the range $4 - 5d$ (see table \ref{table0}).
Adopting $n_0/\Omega_s=1,$ and scaling to the observed orbital period, we find  after setting both $\cos\beta$
 and $\sin\beta$ to equal unity in  (\ref{Gaussequation4.3z2a}) that
$t_{\beta}$  exceeds  $\sim 4.7\times 10^{13} y.$
Here and unless otherwise stated in this Section, for simplicity we assume $M_p=1$ Jupiter mass and take $M_s$ and $R_s$ to be the same as for our $1.3M_{\odot}$ stellar model.

 For HAT-P-9 the system parameters are approximately the same apart from the orbital period being somewhat smaller at $3.92d.$
From  the observed $V\sin i$ with $i=90^{\circ}$ we estimate a rotation period of $\sim 5.6d.$\footnote{From (\ref{Gaussequation4.3z2a}) and  (\ref{Gaussequation4.3z2})
it follows that reducing the rotation period on account of $i < 90^{\circ}$ increases the bound on $t_{\beta}.$}
Thus adopting one Jupiter mass for $M_p$  from  (\ref{Gaussequation4.3z2a}) and  (\ref{Gaussequation4.3z2})
we find  the lower limit for $t_{\beta} $ is smaller by  a factor $\sim 3$ in this case as compared to the value for WASP-7.

The parameters for the central stars in  Kepler-8 and KELT-4 are similar to those of WASP-7  (see table \ref{table0}).
However the orbital periods are $3.52d$ and $2.99d$ respectively and the rotation periods estimated as above from $V\sin i$ are $6.25d$
and $14d$ respectively. 
The corresponding bounds on $t_{\beta}$ are $\sim  7.8\times 10^{12}y$ and $\sim 1.9\times 10^{12}y$  respectively.

Thus tides are not expected to produce significant realignment in any of these cases and are  less
effective by a factor exceeding  between $48$ and $1200$ than is implied by the estimate for a solar mass star given by PS1 ( see the discussion below their equation (47)).
However, we remark that some of this variation occurs  because
tidal evolution times are shortened as the rotation period becomes longer, as well as when the orbital period becomes shorter, as was indicated by PS1 
when considering  the case
of a solar mass star.

To consider the possibility of slow stellar rotation consider HAT-P-30. This system has an orbital period of  $2.81d.$  
The small value of $V\sin i$ for $i=90^{\circ}$  indicates an upper bound for the stellar rotation period in the range $25-40 d.$ 
Adopting the value $25d$  leads to  $t_{\beta} > \sim 2.6\times 10^{11}y.$ This  indicates the possibility of beginning to approach a regime where tides may start to become significant.
However, this system shows strong misalignment. For $\beta = 80^{\circ},$
 the factor $(1+L/S\cos\beta)$ in
(\ref{Gaussequation4.3z2a}) is reduced by a factor $\sim 6$ compared to when $\beta=0,$ as was assumed in that factor when making the above estimate.
It should accordingly be increased by the same factor.

\subsection{Short period systems}\label{shortperiod}
We can also extend the discussion to 
consider short period systems such as WASP-12 and WASP~-~18 which have similar central stars but orbital periods $ 1.09d$ and $0.94d$
respectively. Equations  (\ref{Gaussequation4.3z2a}) and  (\ref{Gaussequation4.3z2}) for an orbital period of $0.94d$,  an adopted 
stellar rotation period of $6d,$ $M_p=10$ Jupiter masses and $\beta=\pi/2$ yields for a system resembling WASP-18 that
$t_{\beta} >\sim  9\times 10^{9}y.$

 Notably the orbital period of WASP-12 which has a strong spin-orbit misalignment has been observed to decay on a $10^6 y $ time scale 
\citep[][]{Yee}. This implies that tidal forcing terms with $|m|=2$  
which the above discussion has  neglected may be important for the evolution of very short period systems.
However,   orbital decay  has not yet \textcolor{majenta}{been} found for the aligned system WASP-18 discussed above.

\section{Summary and discussion}\label{discuss}
In Section \ref{tidalresponse} we summarised the procedure used for obtaining the tidal response  to forcing 
of a $1.3M_{\odot}$  star by  potential perturbations with angular  dependence \textcolor{majenta}{proportional} to a spherical  harmonic  of degree $2.$
Numerical results were then discussed in Section \ref{NumRes}.
As in our previous work in PS and PS1 we found $r$ modes by searching for clear  resonant responses to tidal forcing.
Their main properties were discussed in Section \ref{rmoderes}. Results for
$r$ modes with $l'=1$ were given in Section \ref{l'=1}
and for $r$ modes with $l'=3$ in Section \ref{l'=3}.
In each of these cases there is a spectrum of modes closely spaced in frequency with increasing number of radial nodes
in the radiative interior as was found in our previous work in PS and PS1 for a solar model.

As for stars that are in the neighbourhood of the Kraft break, or hotter,  the convective envelope is thin,
we found it useful to relate our numerical results to the response of a simplified
thin convective envelope model based on vertical averaging.   This is  summarised in  Section \ref{thinconvaveraging}. and discussed in more detail in
appendices \ref{AS1}-\ref{AS4}.  This approach is similar to that followed when obtaining  the Laplace tidal equations \citep[e.g.][]{LH} for an incompressible ocean.
Initially we allowed for  centrifugal distortion in order to properly discuss the rigid tilt mode.   This was subsequently neglected to correspond to the 
simplified  stellar model for which we obtained numerical results.

In this simplified model  the angular dependence of the response is determined  by a single governing  equation that can be  expressed in the forms (\ref{B4}) - (\ref{B4a}).
The tidal response obtained from consideration  of this model is discussed further in appendices \ref{HM} and \ref{rmodes}  where
the associated  free $r$ modes and their connection
 to Hough eigenfunctions  is considered.
 
The relation
to the rigid tilt mode ( that occurs for $l'=|n|=1$) is  discussed in appendix \ref{RTmode}.
We  further  consider how the convective envelope can participate in a near rigid  tilt mode  when  only that  is involved and thus self gravity, which has to be included in order
to obtain the proper rigid tilt mode in which the whole star participates, is negligible in appendix \ref{Neartilt}.

In addition in appendix \ref{Pertt1a} we  estimated  the departure of the $r$ mode frequencies from
their values in the zero frequency limit and made a  reasonably successful  comparison  with our numerical work in Sections \ref{l1nr0resonanceshift}
and \ref{l'=3}.
This is significant for understanding why  $r$ modes are effectively
non resonant for the  rotation periods of interest.

{ In deriving the simplified model employing vertical averaging, we introduced two small parameters, 
 the ratio of the thickness to radius of the convective envelope, $\epsilon_1$ and
the centrifugal parameter $\epsilon_2$ (see  appendix \ref{AS3}). The latter parameter is also assumed to be small in the numerical treatment in this  and previous papers
(PS and PS1) being always $ < \sim 1.5\times 10^{-3}.$
In addition the oscillation frequency is assumed to be of order $\Omega_s.$
 
 For the stellar model considered in this paper for  which $\epsilon_{1} = 0.12$ we obtain reasonable
  agreement with the simplified model that would also be expected for smaller values, 
( see the discussion in Section \ref{l'=1}).  In particular the ratio of the magnitude of the radial  displacement to the  magnitude of the horizontal displacement
 in the convective envelope is characteristically  of order $\epsilon_2$ with the magnitude of the horizontal displacement being  approximately  independent of radius. 
However, for the solar model considered in PS1
 for which $\epsilon_1= 0.27$  the latter is not the case.
}

We went on to find the non resonant  tidal response of the $1.3 M_{\odot}$ model for rotation periods in the range $5-31$ $d$ in Section \ref{Nonresresults}.
We were thus able to consider short enough rotation periods to correspond to those observed for many hot Jupiter systems
with appropriate central stars, though not large enough to imply significant centrifugal distortion.

In particular we considered responses to  secular tidal forcing with $m=0.$
Such forcing only occurs when the spin and orbital angular momenta are misaligned and in a frame  co-rotating  with the star
appears to have forcing frequencies comparable to $\Omega_s.$ This may lead to the excitation of inertial modes in the
convective envelope and thus an enhanced tidal response \citep[see eg.][]{Albrecht12,Lai12}.
This has been postulated  to  allow alignment to occur without there being orbital decay  that would lead to synchronisation.
That would result in tidally aligned systems which do not manifest orbital decay provided that  the star has a large enough convective envelope.
This process is accordingly expected to be less effective for stars at or beyond the Kraft break as compared to a solar type star.

We discussed the effects on tidal evolution under these circumstances in Section \ref{Effecttidalevol}
presenting expressions for the rate of evolution of the spin orbit angle under the assumption 
of fixed orbital angular momenta in Sections \ref{spinorbevol}.
- \ref{m0evol}.
These were applied to  systems with orbital periods in the range $2.8-5$ days listed 
 in table \ref{table0} in Section \ref{2.8days}.
In all these cases a lower bound on the alignment time scale, obtained under the assumption of fixed orbital angular momentum
indicated very little realignment should occur in system lifetimes. The process is less effective  than for systems with  a solar type star
on account of both weaker dissipation in a smaller convective envelope\footnote{As a comparison with the results in PS1 indicates, the dissipation rates associated with forcing with $(n,m) =(-1,0)$ and $(n,m)=(-2,0)$
are significantly less than what was  found for a solar model in PS1 for the same value of $R_s/a.$}
and  the shorter rotation periods expected for stars at or beyond the Kraft break.

{  According to \citet{Beyer} the Kraft break transition involves a narrow mass range of $1.32 M_{\odot } - 1.41 M_{\odot}$ with a corresponding range in effective temperature of $200K.$ In addition there is a transition from slow to rapid rotation. From \citet{Beyer}
 our calculations are seen to be  for a star at the low mass end of the Kraft break where slowly rotating stars may be observed as indicated in table \ref{table0}.  
 As noted above slow  rotation tends to shorten alignment time scales as compared to rapid rotation on account of the relatively lower stellar angular momentum
 content ( see discussion in Section \ref {m0evol}).  As passing through the Kraft break is associated with the disappearance of the convective envelope the tidal dissipation rate significantly reduces. As noted above in Section \ref {2.8days}, already for the $1.3 M_{\odot}$  model
   the tidal realignment time scale is increased by a factor in the range $48-1000$ when compared to a solar mass model.
Without a convective envelope we estimate that reliance on radiative dissipation would increase this even further by up to two orders of magnitude.}

As indicated in Section \ref{shortperiod}{, different considerations may be required} for systems with very short orbital periods $<  \sim 1d$
such as WASP-12 and WASP-18. As exemplified by WASP-12 such systems may be undergoing orbital decay on a significantly shorter time scale
than  that estimated here for realignment  \citep[][]{Yee}. 
Rapid orbital decay may be associated with the increasing importance of the excitation of inwardly propagating $g$ modes
(see Section \ref{-2,-2})  as the orbital period decreases. 
In  this case any realignment  is likely connected to that process.

Finally. we stress  the limitations of our results  on account of the simplified model in which a spherical star without centrifugal
distortion was adopted. This limits the rotation  periods that can be considered and may affect the details of the $r$ mode response.
This should be considered in future work.

\section{Data Availability}
The data underlying this article will be shared on reasonable request to the corresponding author.



\begin{appendix}
\textcolor{majenta}{
\section{The response of a  convective envelope in a thin shell obtained by making  use of a vertical averaging approximation}\label{AS1}
We find it useful to allow for the possibility that the boundaries of the shell are not spherical as this allows centrifugal distortion to be discussed
and the effect of ignoring this clarified. It turns out that its neglect does not produce qualitative changes
  when self-gravity can be neglected, though inclusion of both self-gravity and centrifugal distortion, neglected in our numerical modelling,
   is essential for correctly  incorporating the rigid tilt mode.
Though incorporating the latter,    which is not   directly  associated with dissipation,  is not important for the  tidal evolution we consider (see PS1). }

\textcolor{majenta}
{To allow for the possibly of centrifugal distortion  we introduce a general  orthogonal coordinate system with axial symmetry, $(\psi,\chi,\phi)$, which becomes a spherical polar   
coordinate system when $\psi \rightarrow r,$ and $\chi \rightarrow \theta$ (see \citet{PP78} who worked with such a system).
The surfaces defined by $\psi= {\rm constant}$ are equipotential surfaces and  for an assumed barotropic equation of state, in hydrostatic equilibrium the pressure and density are such that
 $P=P(\psi)$ and $\rho=\rho(\psi).$
 Thus the unit vector,
  ${\hat{\bsf {\psi}}} = \nabla\psi/ |\nabla\psi |,$
   is normal to a surface of constant pressure and density.}
   
     \textcolor{majenta}{ A good approximation for the surfaces $\psi =$ constant for centrally condensed stars such as those we consider 
   is obtained from the Roche  potential  which applies in the limit of high central condensation of the primary and 
    for which the equipotential surfaces are given by \citep[see e.g.][]{PW} 
   \begin{align}
   \frac{GM}{\psi}=\frac{GM}{r}+\frac{1}{2}r^2\Omega^{2}_s\sin^2\theta\label{ROCHE}.
   \end{align}
 Thus $\psi$ is the radius at the pole $\theta=0.$ The orthogonal $\chi$ coordinate is given to first order in $\Omega_{s}^2/\Omega_{c}^2$ by
 \begin{align}
 \chi=\theta + \frac{\Omega_{s}^2 r^3\sin\theta\cos\theta}{3GM_*}\label{ROCHE1}
 \end{align}
 }

 \textcolor{majenta}
{In this coordinate system  the angular velocity vector ${\bsf \Omega}_s = (\Omega_{s,\psi}, \Omega_{s,\chi},0 ),$
where
\begin{align}
 &\hspace{-1.3cm}
  \Omega_{s,\psi} = \Omega_s {\hat {\bf k}}\cdot  {\hat {\bsf \psi}},\; {\rm and}\;  \Omega_{s,\chi} = \Omega_s {\hat {\bf k}}\cdot  {\hat{\bsf \chi}},\;
 {\rm  with}\; {\hat {\bf k}},\; {\rm and}\;{\hat{\bsf \psi}}=\nabla\psi/|\nabla\psi |,\; {\rm and}\; 
   {\hat{\bsf \chi}}=\nabla\chi/|\nabla\chi|\nonumber \\
 &\hspace{-1.3cm}
 {\rm being\ unit\ vectors\ in\ the}\
  z, \psi,\; {\rm  and}\; \chi \ {\rm  directions\ respectively.}\nonumber\label{A1}
  \end{align}
Similarly the general Lagrangian displacement  is written
${\bsf \xi}= (\xi_{\psi},\xi_{\chi},\xi_{\phi}).$
\subsection{Governing linearised equations}\label{AS2}
Assuming the convective envelope can  be modelled by  an inviscid  barotropic fluid,
denoting perturbation quantities with a prime,  and assuming time dependence through a factor $\exp({\rm i}\sigma t)$,  the components of the linearised equation of motion are
\begin{align}
&-\sigma^2\xi_{\psi} +2{\rm i}\sigma\Omega_{s,\chi}\xi_{\phi}=-|\nabla \psi | \frac{\partial {\cal W}'}{\partial \psi}\nonumber\\
&-\sigma^2\xi_{\chi} - 2{\rm i}\Omega_{s,\psi}\xi_{\phi}=-|\nabla \chi | \frac{\partial {\cal W}'}{\partial \chi}\nonumber\\
&-\sigma^2\xi_{\phi} + 2{\rm i}\sigma(\Omega_{s,\psi}\xi_{\chi} -\Omega_{s,\chi}\xi_{\psi})   =-\frac{1}{r\sin\theta} \frac{\partial {\cal W}'}{\partial \phi},\;{\rm where}
\;{\cal W}'=\frac{P'}{\rho}+\Phi' = \frac{\rho'c^2}{\rho}+\Phi',
\end{align}
where $\Phi'$ is the sum of the  perturbation to the gravitational  potential  and the forcing tidal potential and $c$ is the sound speed.
The oscillation frequency $\sigma,$ assumed to be of order $\Omega_s,$  may be free or forced in the latter case $\sigma\equiv \omega_{f,n,m}.$ 
The linearised continuity equation leads to
\begin{align}
\frac{ \rho({\cal W}'-\Phi')}{c^2}=-\frac{|\nabla \chi | |\nabla \psi | }{r\sin\theta} \left( \frac{\partial }{\partial \psi} \left(\frac{\rho r\sin\theta \xi_{\psi}}{|\nabla \chi | }  \right)+
 \frac{\partial }{\partial \chi} \left(\frac{\rho r\sin\theta \xi_{\chi}}{|\nabla \psi| }  \right)  +\frac{\rho}{|\nabla \chi | |\nabla \psi |}\frac{\partial\xi_{\phi}}{\partial\phi}        \right)\label{A2}
\end{align}
\subsection{Relevant small parameters}\label{AS3}
There are two small parameters we consider in this analysis, the characteristic ratio of the  pressure scale height  to the radius of the convective envelope,
 $\epsilon_1= h/R_s$ and the centrifugal parameter, $\epsilon_2=\Omega_s^2R_s^3/(GM_s).$
 Here we assume that the total thickness of the layer is not significantly less than the local  pressure scale height at any interior point and comparable to that at the lower boundary,
 thus this could be taken to be $h.$
We adopt a scheme in which formally $\epsilon_2\ll \epsilon_1\ll 1.$  For a specified, $h/R_s,$ this can always be arranged for sufficiently small $\Omega_s.$ }

\textcolor{majenta}{In the inertial regime  for which  $|\sigma| < 2 \, \Omega_s$ we expect,  as will  be verified a posteriori,  that for the unforced  free oscillation case with $\sigma$ being of the same order as $\Omega_s$ and for which $\Phi'=0,$  the characteristic  ratio of the  vertical and horizontal displacements, $|\xi_{\psi}|/|\xi_{\perp}|= O(\epsilon_2),$ with  $|\xi_{\perp}|$  taken to be the magnitude of the horizontal component of the displacement.\footnote{\textcolor{majenta}{In the case with forcing this condition  applies to the deviation 
 of ${\bsf \xi}$ 
 from a suitably defined  equilibrium tide.}} The  case where $\sigma=\pm2\Omega_s,$ corresponding to the boundary of the inertial regime, is a special case which will be discussed further below.}
 
 In addition from the second and third components of  (\ref{A1}),  characteristically $|{\cal W'}|/|g \xi_{\perp}|=O(\epsilon_2),$ 
 where $g = \left | (|\nabla \psi |/\rho)dP/d\psi\right|$ 
is the local gravitational acceleration. 
It then follows that the term  $\propto {\cal W}'$ on the left hand side of (\ref{A2}) 
  is  characteristically smaller in magnitude than  each of the  terms enclosed in the outermost brackets on the right hand side by a factor $\epsilon_2/\epsilon_1\ll 1.$ However, this is $ \gg \epsilon_2,$  which measures the fractional correction to each of  these  terms potentially arising from centrifugal distortion.
\textcolor{majenta}{ That is  because the  parameter $\epsilon_2$  measures  the departures  
of $\psi$ from $r$, $\chi$ from $\theta,$ as well as $|\nabla\psi|$ and $r|\nabla\chi|$ from unity.}

  Similarly, corrections due to centrifugal distortion in the second and third  equations of the set  (\ref{A1}) are smaller by a factor $\epsilon_2$ than the terms that dominate. 
   This also applies to the term $\propto \xi_{\psi}\Omega_{\chi}$  in the third equation listed in (\ref{A1}) and so this will  be neglected  from now on.
   
   \textcolor{majenta}{\subsubsection{The special case with $\sigma^2= 4\Omega_s^2$}\label{epicyclic}
    At this point we recall that as noted above, the preceding discussion does not apply to the case where $\sigma=\pm 2\Omega_s.$
   This is because when  the system (\ref{A1}) is regarded as a system of linear equations for the determination of the components
   of $\bmth{\xi},$} in terms of ${\cal W}',$  it is necessary to invert a singular matrix when $\sigma^2=4\Omega_s^2.$ This is because of the
   existence of free epicyclic oscillations when ${\cal W}'=0.$ This situation implies a constraint on ${\cal W'}$ or the forces acting required for solubility.
   In addition one can add a free epicyclic oscillation with arbitrary amplitude which needs to be determined through further consideration of the full solution.
   Viscosity should also be included. As from (\ref{A1}) an epicyclic oscillation has, $\xi_{\psi}=\pm {\rm i}{\hat {\bf k}}\cdot  {\hat{\bsf \chi}} \xi_{\phi},$ we cannot \textcolor{majenta}{conclude that the magnitude of the vertical component is very much smaller than that of the horizontal components. }
   \footnote{If viscosity is ignored one  concludes that ${\cal W}'=0$ and only an epicyclic oscillation is present, the amplitude being determined from consideration of the linearised continuity equation.}

\textcolor{majenta}{
\subsection{Hydrostatic equilibrium in the vertical direction and vertical averaging}\label{AS4}
\textcolor{majenta}{Because the convective  layer is thin, we may infer from the first equation of the set (\ref{A1})  that, after neglecting centrifugal distortion, $\partial{\cal W}'/\partial \psi~=~0$ to within a relative  correction of order $\epsilon_1.$ }
Thus ${\cal W}'$ is a function of $\chi$ and $\phi$ alone corresponding to an assumption of hydrostatic equilibrium in the vertical direction.
Using this it then follows from the second of the set of equations  (\ref{A1}) that the same is true for the horizontal components of the Lagrangian displacement to within a relative error of order $\epsilon_2.$
The form of ${\cal W}'=P'/\rho +\Phi'$ can be found by evaluating it at the surface boundary where we assume the Lagrangian pressure perturbation to vanish or $P' = -\xi_{\psi}|\nabla\psi|dP/d\psi .$
Thus
 \begin{align}
 {\cal W}' = \xi_{\psi,R_s}g+\Phi' .\label{B1}
 \end{align}
  Here  $\xi_{\psi,R_s}$   is  $\xi_{\psi}$  evaluated on the surface boundary 
(assumed to be an equipotential) as is the gravitational acceleration, $g.$ We remark that the latter is approximately constant  in a thin convective  layer. }

\textcolor{majenta} { It can also be seen from (\ref{B1}) and the estimate from (\ref{A1})  giving
  $|W'|\sim  R_s\Omega_s^2|\xi_{\perp}| $\footnote{\textcolor{majenta}{Note that this estimate assumes that $\sigma$ is comparable to $\Omega_s$ When $\sigma$ is significantly smaller
  as for high $l'$ $r$ modes it should be correspondingly reduced}}  that
  in the free oscillation case with negligible self-gravity  with  $\Phi'=0,$ $|\xi_{\psi}/\xi_{\chi}|= O(\epsilon_2)$ as assumed  above.
  When $\Phi' \ne 0$ this ordering applies to $|\xi_{\psi}+\Phi'/g|,$ noting that $-\Phi'/g$ corresponds to the equilibrium tide vertical displacement at the surface boundary.}

\textcolor{majenta} {From the above considerations, at the lowest order ${\cal W}',$  $\xi_{\chi}$ and $\xi_{\phi}$ 
depend only on $\chi$ and $\phi.$  This  enables us to implement the procedure  of vertical integration/averaging adopted in the derivation. of the Laplace tidal equations
\citep[see e.g.][]{LH}. 
We define the vertical average of a quantity, $Q,$ as \footnote{ \textcolor{majenta}{$R_s$ and $R_{ce}$ here define the surface and lower boundary equipotential surfaces and will be approximately equal to the radii
of any points on them to within a relative error of order $\epsilon_2$. For convenience these radii can be evaluated at $\theta=\chi=0.$}}
\begin{align}
\langle Q\rangle= \frac{\int^{R_s}_{R_{ce}}\rho Qd\psi}{\Sigma} \;{\rm where\ the\ surface\ density};\Sigma =  \int^{R_s}_{R_{ce}}\rho d\psi\;{\rm is\ constant.}
\end{align} 
At the lowest order, $\xi_{\chi}, \xi_{\phi}, $ and, ${\cal W}',$ are assumed not to 
depend on $\psi.$ This means they are  equal to  their vertical averages. Accordingly the angled bracket notation for these quantities will be dropped henceforth.  
We also remark that    the quantities, $ r$,  where it appears explicitly, $\sin \theta, |\nabla\psi|,$ and $ |\nabla\chi|,$  expressed as functions of $\psi$ and $\chi,$ can be 
evaluated at any point  on the lower boundary equipotential, which we will take to be   at $\theta=\chi=0,$
with  relative  corrections  of order, $\epsilon_2\epsilon_1,$ which is sufficient for our purposes. 
Proceeding as outlined above and  vertically averaging  the second and third equations of (\ref{A1}) we obtain\
\begin{align}
&-\sigma^2\xi_{\chi} - 2{\rm i}\Omega_{s,\psi}\xi_{\phi}=-|\nabla \chi | \frac{\partial {\cal W}'}{\partial \chi}\nonumber\\
&-\sigma^2\xi_{\phi} + 2{\rm i}\sigma\Omega_{s,\psi}\xi_{\chi}    =-\frac{1}{R_{ce}\sin\theta} \frac{\partial {\cal W}'}{\partial \phi},\label{B2}
\end{align}
\begin{align}
&{\rm which\; lead\; to\; }\quad  (\sigma^2- 4\Omega_{s,\psi}^2)\xi_{\chi}= |\nabla \chi | \frac{\partial {\cal W}'}{\partial \chi} - 
\frac{2{\rm i}\Omega_{s,\psi}}{\sigma R_{ce}\sin\theta} \frac{\partial {\cal W}'}{\partial \phi} \nonumber\\
&{\rm and\;}\quad \quad\quad \quad\quad  (\sigma^2- 4\Omega_{s,\psi}^2)\xi_{\phi}    =\frac{1}{R_{ce}\sin\theta} \frac{\partial {\cal W}'}{\partial \phi}
+\frac{2{\rm i}\Omega_{s,\psi} |\nabla \chi |}{\sigma} \frac{\partial {\cal W}'}{\partial \chi} .\label{B21}
\end{align}}
\textcolor{majenta}{Similarly vertically averaging  (\ref{A2}) yields
\begin{align}
&\frac{ \rho_{Rce}|\nabla\psi| ({\cal W}'-\Phi')}{g}\equiv \rho_{Rce} |\nabla \psi |\xi_{\psi,R_s} =  \nonumber \\ 
& \rho_{Rce} |\nabla \psi |\xi_{\psi,Rce} -
 \frac{|\nabla \chi | |\nabla \psi | }{R_{ce}\sin\theta} \left( 
 \frac{\partial }{\partial \chi} \left(\frac{\Sigma R_{ce}\sin\theta \xi_{\chi}}{|\nabla \psi| }  \right)  +\frac{\Sigma}{|\nabla \chi | |\nabla \psi |}\frac{\partial\xi_{\phi}}{\partial\phi}        \right)\label{B3}
\end{align}
where $\rho_{Rce},$ and $\xi_{\psi,Rce}$ are the density and vertical displacement evaluated at the convective envelope inner boundary, and  we have made use of (\ref{B1}) together with
 hydrostatic equilibrium of the unperturbed state in the form $c^2  |\nabla \psi | d\rho/d\psi=-\rho g$ and as above evaluated $g$ on the surface equipotential.
We  also recall that to the accuracy we are working the evaluation can be taken at $\theta =\chi=0.$ 
}

\textcolor{majenta}{Using (\ref{B21}) to eliminate the horizontal displacements in favour of ${\cal W}',$ (\ref{B3}) becomes an equation for ${\cal W}'$ alone that can be written in the generic form
\begin{align}
-\frac{ \rho_{Rce}R_{ce}^2\sigma^2|\nabla\psi|}{\Sigma}\left(\xi_{\psi,R_s} -\xi_{\psi,Rce}
\right)= {\cal L}(W')
\label{B4}
\end{align}
where the operator ${\cal L}({\cal W}'),$ being readily constructed with the help of (\ref{B21}). It  is specified  in the limit of zero centrifugal distortion
for which $\psi \rightarrow r,$ $\chi\rightarrow \theta$ and $\nabla\psi =1 $
at the beginning of  Section \ref{HM} below.
 }
 \textcolor{majenta}{
  In this limit $\Omega_{\psi}\rightarrow\Omega_r = \Omega_s\cos\theta$ while $\Omega_{\chi} \rightarrow\Omega_{\theta}$ no longer
  appears as in the traditional approximation. In addition,  as the layer is thin, where $r$ appears explicitly it is replaced by $R_{ce}.$ }
  
 \textcolor{majenta}{ However, we note that 
   as $\psi \rightarrow r,$ it is a coordinate that is constant on total equipotential surfaces that are close to spheres (see e.g. \ref{ROCHE}), deviating by an amount measured by $\epsilon_2.$
  This means that  the direction of $\xi_{\psi}$ deviates from the purely radial direction by an angle of order $\epsilon_2.$ 
  Because the layer is thin we then have $\xi_{\psi,R_s}-\xi_{\psi,Rce}\rightarrow (\bmth {\xi}_{R_s}-\bmth{\xi}_{Rce})\cdot{\hat{\bf r}}$ with a relative deviation of order $\xi_{\theta} \epsilon_2\epsilon_1$ which is
   expected to be of order ${\cal W}'\epsilon_1/g .$  Thus  in the limit with a relative deviation of order $\epsilon_2$  we have
   \begin{align}
-\frac{ \rho_{Rce}R_{ce}^2\sigma^2}{\Sigma}\left(   \bmth {\xi}_{R_s}-\bmth{\xi}_{Rce}\right )\cdot{\hat{\bf r}}   
= {\cal L}(W')
\label{B4X}
\end{align}
we may also write this as
 \begin{align}
-\frac{ \rho_{Rce}R_{ce}^2\sigma^2}{g\Sigma}\left({\cal W}' - \Phi'- g\xi_{r,Rce}\right) 
= {\cal L}(W'),
\label{B4a}
\end{align}
where we have  used (\ref{B1}) for ${\cal W}'$ with $\xi_{\psi,R_s} \rightarrow \xi_{r,R_s}$ and  $\xi_{\psi,Rce} \rightarrow \xi_{r,Rce}$.
However,  in spite of this re-labelling,  we should still regard these quantities as pointing in the local vertical direction, this
being inclined at an angle $O(\epsilon_2)$ with respect to the local radius. This is a small angle that is neglected in the numerical modelling of this paper.   
 }

 \textcolor{majenta}{ \section{ The tidal response  and the excitation of ${\rm \lowercase{ r}}$ modes }\label{HM}
 For fixed real, $\sigma,$  an assumed $\phi$ dependence through a factor $\exp({\rm i}n\phi),$ the neglect of centrifugal distortion and $\chi\rightarrow\theta,$
 the operator ${\cal L}$ can be seen to be defined through
 \begin{align}
& {\cal L}({\cal W}') = \frac{1}{\sin\theta}\frac{d}{d\theta} \left(\frac{\sin\theta\left( \frac{d{\cal W'}}{d\theta}+\frac{2n\Omega_s\cos\theta {\cal W'}}{\sigma\sin\theta}\right)}
{\left(1-\frac{4\Omega^2\cos^2\theta}{\sigma^2}\right)}\right) -\frac{2n\Omega_s\cos\theta}{\sigma\sin\theta} \left(\frac{\left( \frac{d{\cal W}'}{d\theta}+\frac{2n\Omega_s\cos\theta {\cal W}'}{\sigma\sin\theta}\right)}{\left(1-\frac{4\Omega^2\cos^2\theta}{\sigma^2}\right)}\right)
 -\frac{n^2{\cal W}'}{\sin^2\theta}
\label{C0}
 \end{align}
  Hough functions are defined by the  eigenvalue problem, ${\cal L}({\cal W})=-\lambda {\cal W},$ with eigenvalue $\lambda (\sigma)$ which depends on $\sigma$ \citep[see e.g.][]{SP97, PS23}. 
  The eigenvalue problem  is more conveniently solved through
  considering the pair of equations\footnote{\textcolor{majenta}{This removes the appearance of the  apparent singularity where, $\sigma^2=4\Omega_s^2\cos^2\theta,$
  when  the second equation in (\ref{C1}) is used to eliminate ${\cal Q}$ and thus form ${\cal L}({\cal W}').$ }}
 \begin{align}
& {\cal O}({\cal W}',{\cal Q})
\equiv \frac{1}{\sin\theta}\frac{d({\cal Q}\sin\theta)}{d\theta} -\frac{2n\Omega_s\cos\theta {\cal Q}}{\sigma\sin\theta} -\frac{n^2{\cal W}'}{\sin^2\theta}=-\lambda {\cal W}'\quad {\rm and}\nonumber\\
&{\cal P}({\cal W}',{\cal Q})\equiv\frac{d{\cal W}'}{d\theta}+\frac{2n\Omega_s\cos\theta {\cal W}'}{\sigma\sin\theta}-\left(1-\frac{4\Omega^2\cos^2\theta}{\sigma^2}\right){\cal Q}=0\label{C1}
 \end{align}
We remark that for  ${\cal W}', Q, {\cal W}'_1, Q_1,$ that are regular for $\theta \in [0,\pi],$   the operators  ${\cal O}$ and  ${\cal P}$ satisfy the relations
\begin{align}
&\int^{\pi}_0\sin\theta ({\cal W}'_1)^{*} {\cal O}({\cal W}',{\cal Q})d\theta           
- \int^{\pi}_0\sin\theta  Q_1^* {\cal P}({\cal W}',{\cal Q})d\theta=
 \nonumber\\               
&\int^{\pi}_0\sin\theta {\cal W}' {\cal O}(({\cal W}'_1)^{*},{\cal Q}_1^*)d\theta           
- \int^{\pi}_0\sin\theta Q_1{\cal P}(({\cal W}'_1)^{*},{\cal Q}_1^*)d\theta \label{C2}
\end{align}
from which the orthogonality condition for eigenfunctions ${\cal W}'_i$ and ${\cal W}'_j $  with different eigenvalues $\lambda_i,$ and $\lambda_j,$  $\int^{\pi}_0 ({\cal W}'_i)^* {\cal W}_j\sin\theta d\theta=0,$ follows
as does the fact that ${\cal L}$ is self-adjoint.
We remark that from the form of, ${\cal L}$, each eigenvalue, $\lambda_i$, necessarily real,  is a function of $\sigma/\Omega_s.$}
\textcolor{majenta}{
\subsection{Eigenvalues corresponding to $r$ modes}\label{rmodes}
For the specific values of $\sigma=\sigma_0= 2n\Omega_s/(l'(l'+1),$ $l'$ being an integer $>|n|,$  there is an  eigenvalue such that, $\lambda =0,$ This corresponds to an  $r$ mode with specified, $(l',n),$
in the limit $\Omega_s\rightarrow 0$ \citep[see e.g.][]{ PP78, SP97,PS23}. The corresponding eigenfunction, to within an arbitrary normalising factor is given by
\begin{align}
&{\cal Q}=(1-\mu^2)^{-1/2}P^{|n|}_{l'}(\mu),\quad{\rm and} \nonumber\\
&{\cal W}'= {\cal W}_0 = -\frac{1}{n^2}\left( (1-\mu^2)\frac{d P^{|n|}_{l'}(\mu)}{d\mu}+ l'(l'+1)\mu P_{l}'^{|n|}(\mu)\right), \quad {\rm where}\; \mu=\cos\theta \label{C3}
\end{align}
and  $P^k_{l'},$ denotes the associated Legendre function. We note that the above quantities are real. Furthermore in this limit spherical shells are decoupled 
from each other. An issue that is resolved when the strict zero rotation frequency  limit is departed  from \citep[see][]{PP78,  PS23, PS24}. 
\\
\subsubsection{The rigid tilt mode}\label{RTmode}
In an approximation  scheme that  ignores centrifugal distortion and self-gravity.
 we do not expect to recover the rigid tilt mode. This mode has  $l'=|n|=1$ and accordingly from (\ref{C3}), $ {\cal Q} = {\cal Q}_0= const.,$ and
${\cal W}'=-{\cal Q}_0\sin\theta \cos\theta \propto P^{1}_2(\mu).$ In addition  $\xi_{\theta}={\cal Q}_0/(R_{ce}\Omega_s^2),$ being constant, $\xi_{\phi}= in\xi_{\theta}\cos\theta $
and $\bmth{\xi}\cdot{\hat{\bf r}}=0.$
It is readily verified that this is satisfied by the form of the governing equation (\ref{B4X}).}

\textcolor{majenta} {However, in order for equation (\ref{B1}) to be satisfied the effect of self-gravity has to be included. If we write $\Phi'=\Phi'_{SG}$ where $\Phi'_{SG}$ is the perturbation of the potential
arising from self-gravity, $\Phi_{SG}.$ Then as for the rigid tilt mode the Lagrangian perturbation to $\Phi_{SG}$ is zero, \\
we have $\Phi'_{SG}= -\bmth{\xi}\cdot \nabla \Phi_{SG}.$ Inserting this into (\ref{B1}) we obtain\\
${\cal W}' =-(1/\rho)\bmth{\xi}\cdot \nabla P-\bmth{\xi}\cdot\nabla \Phi_{SG}=\bmth{\xi}\cdot\nabla \Phi_{C}= -\xi_{\theta}\Omega_s^2 R_s\sin\theta\cos\theta .$\\
 Here $\Phi_C= -(\Omega_s^2/2)r^2\sin^2\theta$ is the centrifugal potential, and we have made use of hydrostatic equilibrium while
  recalling that  ${\cal W}'$ is evaluated on the surface  $r=R_s.$ Thie above expression  coincides with the form of ${\cal W}'$ specified
 at the beginning of this Section.}
 
\textcolor{majenta}{ \subsubsection{Near tilt modes in the convective envelope}\label{Neartilt}
As we do not include self-gravity  in our modelling there will be no rigid tilt mode. This would be the situation where the disturbance in the radiative interior is short wave length making
$\Phi'_{SG}$ negligible. Then, as we indicate below  there are solutions in the convective envelope for which ${\cal W}'$ is close to  the form for the rigid tilt mode.
However, in this case, instead of ${\cal W}'$ being balanced by the effect self-gravity,
 it is instead balanced by the introduction of a small but non zero radial displacement on the order of $\epsilon_2\xi_{\theta}.$
 Thus equation (\ref{B1}), with $\Phi'=0,$ applies, but in this case specifying $\xi_{\psi}.$
  This enables the convective envelope to approximately undergo a rigid tilt while
the interior does not participate.
\subsection{Eigenvalues for nearby  frequencies obtained using perturbation theory} \label{Pertt}
The eigenfunctions   associated with the eigenvalue problem ${\cal L}({\cal W})=-\lambda{\cal W}$, may be taken to be real. We can then find the change in $\lambda,$ $\equiv \delta\lambda,$ corresponding to a change $\delta \sigma$ to $\sigma_0.$ from perturbation theory.
Use of the symmetry property given by (\ref{C2}) results in the following expression for, $\delta \lambda,$ that does not depend on the perturbation to the eigenfunction, being
\begin{align}
-\delta\lambda\int^{1}_{-1}{\cal W}_0^2d\mu=\frac{\delta\sigma}{\sigma_0}\int^{1}_{-1}\left(  \frac{2l'(l'+1)\mu Q{\cal W}_0}{\sqrt{1-\mu^2}}+\frac{2\mu^2(l'(l'+1))^2Q^2}{n^2} \right)d\mu\label{C4}
\end{align}
When the eigenfunction given by  (\ref{C3}) corresponding to $\lambda=0,$ , we find  from (\ref{C4}) that for, $l'=-n=1,$  $\delta\lambda = -10\delta\sigma/\sigma_0 \equiv \delta\lambda_{1,-1}(\sigma).$ For, $l'=3,n=-2,$
we obtain $\delta\lambda=- 12\delta\sigma/(13\sigma_0)=- 0.93\delta\sigma/\sigma_0 \equiv \delta\lambda_{3,-2}(\sigma).$ }

\textcolor{majenta}{
\subsection{Eigenvalues for free $r$  modes when  $\Omega_s$ is finite but small and self-gravity is neglected}\label{Pertt1a}
In this case the governing equation is (\ref{B4a}) with $\Phi'$ being eventually set to zero. We rewrite this equation in  the form
\begin{align}
-\frac{ \rho_{Rce}R_{ce}^2\sigma^2}{g\Sigma}\left({\cal W}' - \Phi'- g\xi_{r,Rce}(W',\Phi',\sigma)\right)= {\cal L}(W'),
\label{B4an}
\end{align}
where we recall that here and below $g,$ is to be evaluated at the stellar surface.}

\textcolor{majenta}{We assume that for a given, $\sigma,$ $\xi_{r,Rce}$ is determined by the response of the radiative interior to  $W'$ and $\Phi'$ at the inner boundary of the convection zone.
 These provide a surface boundary  condition there for determining the inner response. 
Accordingly we write 
\begin{align}
 \xi_{r,Rce}\equiv\xi_{r,Rce}({\cal W}',\Phi',\sigma)= \xi_{r,Rce}({\cal W}',0,\sigma)+\xi_{r,Rce}(0,\Phi',\sigma)\label{forcesplit}
 \end{align}
as a linear operator acting on $W'$ and $\Phi'.$\footnote{ \textcolor{majenta}{The last equality ensures linearity as a sum of linear operators and that $\xi_{\psi,Rce}(0,0,\sigma)=0.$ }}}

\textcolor{majenta}{The parameter, $\rho_{Rce}R_{ce}^2\sigma^2/(g\Sigma),$ on the left hand side is of order $,\epsilon_2/\epsilon_1,$
which is taken to be  small.  If this is set to zero an eigenfunction exists corresponding to an $r$ mode with 
$\sigma=\sigma_0= 2n\Omega_s/(l'(l'+1)).$ The corresponding eigenfunction
is the Hough mode with, $\lambda=0,$\\
 ( see appendix \ref{HM} and Section \ref{rmodes}). The eigenfunction is specified by (\ref{C3}). In that case we have ${\cal W}' ={\cal W}_0.$}

\textcolor{majenta}{The correction to $\sigma$ for small finite,  $\rho_{Rce}R_{ce}^2\sigma^2/(g\Sigma),$ is readily found from perturbation theory with the help of results given in Sections \ref{HM}
and  \ref{Pertt}.
Setting $\sigma=\sigma_0+\delta\sigma,$ multiplying (\ref{B4an}) by ${\cal W}_{\sigma}\sin\theta,$ 
 and integrating over $\theta$ we obtain correct to first order
  \footnote{\textcolor{majenta}{ The right hand side  follows from considering the change of Hough eigenvalue $\lambda$ resulting from the change $\delta\sigma$ noting that  that  ${\cal L}$ is self-adjoint }}
\begin{equation}
\frac{ \rho_{Rce}R_{ce}^2\sigma_0^2}{g\Sigma}\int^{\pi}_0\sin\theta {\cal W}_0\left( {\cal W}_0 -g\xi_{r,Rce}({\cal W}_0 ,0, \sigma)\right)d\theta
= \delta\lambda_{l',n}(\sigma) \int^{\pi}_0{\cal W}_0^2\sin\theta d\theta,  \label{D1}    
\end{equation}
where $\delta\lambda_{l',n}(\sigma)$ is calculated in section \ref{Pertt} (see equation (\ref{C4})).
Note that we have retained the full value of $\sigma$ in   the expression, $\xi_{r,Rce}({\cal W}_0,\sigma),$ that occurs in (\ref{D1}). 
This is because although the effects of the corresponding term  are  expected to be small in magnitude, it may vary rapidly with $\sigma,$
e.g. exhibiting strong peaks at characteristic values, 
 leading to the possibility
of a series of split modes of different radial order as considered  by \citet{PS23}. 
Here we undertake only a qualitative discussion of this aspect.
\subsection{Response to tidal forcing at frequencies close to an $r$ mode frequency}\label{Tidalresponse}
To take tidal forcing into account we include the perturbing tidal potential by setting it equal to $\Phi'$ in (\ref{B4a}) which  we recall as
\begin{align}
-\frac{ \rho_{Rce}R_{ce}^2\sigma^2}{g\Sigma}\left({\cal W}' -\Phi' -g\xi_{r,Rce}(W',\Phi',\sigma)\right)= {\cal L}(W')
\label{D2}
\end{align}
or making use of (\ref{forcesplit})
\begin{align}
-\frac{ \rho_{Rce}R_{ce}^2\sigma^2}{g\Sigma}\left({\cal W}' -\Phi' -g 
(\xi_{r,Rce}({\cal W}',0,\sigma)+\xi_{r,Rce}(0,\Phi',\sigma))
\right)= {\cal L}(W')
\label{D2a}
\end{align}
Here we assume that $\sigma$ is close to $\sigma_0$ so that ${\cal W}'$ for the response will be close to   
  ${\cal W}_0.$
Setting, ${\cal W}'= {\cal E}{\cal W}_{\sigma},$ plus a small correction, we find the constant  amplitude ${\cal E}$ by inserting this in (\ref{D2a})  multiplying by
 ${\cal W}_{\sigma} \sin\theta,$ and integrating over $\theta.$ Separating out the terms involving $\Phi'$  we thus obtain
 \begin{align}
&{\cal E}\left[\frac{ \rho_{Rce}R_{ce}^2\sigma_0^2}{g\Sigma}\int^{\pi}_0\sin\theta {\cal W}_0\left( {\cal W}_0 -g\xi_{r,Rce}({\cal W}
_0,0,\sigma)\right)d\theta- \delta\lambda_{l',n}(\sigma) \int^{\pi}_0{\cal W}_0^2\sin\theta d\theta\right]= \nonumber\\
 & \frac{ \rho_{Rce}R_{ce}^2\sigma_0^2}{g\Sigma}\int^{\pi}_0\sin\theta {\cal W}_0\left(\Phi'  - g\xi_{r,Rce}(0 ,\Phi', \sigma)\right) d\theta ,\label{D3}    
\end{align}
 If equation (\ref{D2}) with $\Phi'$ set to zero is satisfied for $\sigma=\sigma_k $ corresponding to a normal mode,  we may make use of this to rewrite (\ref{D3}) in the form
  \begin{align}
&{\cal E}\int^{\pi}_0\left(  \frac{ \rho_{Rce}R_{ce}^2\sigma_0^2}{\Sigma}( \xi_{r,Rce}({\cal W}
_0, 0,\sigma)-\xi_{r,Rce}({\cal W}_0,0,\sigma_k))
+(\delta \lambda_{l',n}(\sigma)-\delta\lambda_{l',n}(\sigma_k)){\cal W}_0\right) \sin\theta {\cal W}_0d\theta  
 \nonumber\\
 &= - \frac{ \rho_{Rce}R_{ce}^2\sigma_0^2}{g\Sigma}\int^{\pi}_0\sin\theta {\cal W}_0\left(\Phi' - g\xi_{r,Rce}(0 ,\Phi', \sigma)\right) d\theta .\label{D4}    
\end{align}
   As expected a singular response is obtained at resonance.  If the lower boundary displacement terms are neglected, one obtains the simple result
    \begin{align}
&{\cal E}(\delta\lambda_{l',n}(\sigma)-\delta\lambda_{l',n}(\sigma_k))\int^{\pi}_0  
 \sin\theta {\cal W}^2_0d\theta  
 \nonumber\\
 &= - \frac{ \rho_{Rce}R_{ce}^2\sigma_0^2}{g\Sigma}\int^{\pi}_0\sin\theta {\cal W}_0\Phi'd\theta .\label{D5}    
\end{align}
 However, it is important to emphasise that one can only recover a series of modes with eigenfrequencies, $\sigma_k,$
 rather than just a single mode,  if the lower boundary terms are included.
 In this respect if dissipation is included in their contribution, $\sigma_k$ becomes complex. In this case if the forcing frequency $\sigma$
is restricted to be real, the contribution of multiple resonance peaks in the response may be reduced or even largely suppressed.    }

\end{appendix}
\end{document}